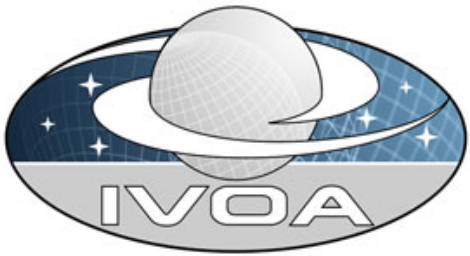

# Describing Simple Data Access Services
# Version 1.0

## IVOA Recommendation 5 October 2013

**This version:**
  http://www.ivoa.net/Documents/SimpleDALRegExt/20131005/
**Latest version:**
  http://www.ivoa.net/Documents/SimpleDALRegExt/
**Previous versions:**
  PR: http://www.ivoa.net/Documents/SimpleDALRegExt/20130911/
  PR: http://www.ivoa.net/Documents/SimpleDALRegExt/20121116/
  PR: http://www.ivoa.net/Documents/SimpleDALRegExt/20120517/
  WD: http://www.ivoa.net/Documents/SimpleDALRegExt/20110921/

**Authors:**
  Raymond Plante, Editor,
  Jesus Delago,
  Paul Harrison,
  Doug Tody,
  and the IVOA Registry Working Group

---

## Abstract

An application that queries or consumes descriptions of VO resources must be able to recognize a resource's support for standard IVOA protocols. This specification describes how to describe a service that supports any of the four fundamental data access protocols--Simple Cone Search (SCS), Simple Image Access (SIA), Simple Spectral Access (SSA), Simple Line Access (SLA)--using the VOResource XML encoding standard. A key part of this specification is the set of VOResource XML extension schemas that define new metadata that are specific to those protocols. This document describes in particular rules for describing such services within the context of IVOA Registries and data discovery as well as the VO Standard Interface (VOSI) and service self-description. In particular, this document spells out the essential mark-up needed to identify support for a standard protocol and the base URL required to access the interface that supports that protocol.

## Status of this document

This document has been produced by the IVOA Registry Working Group.

It has been reviewed by IVOA Members and other interested parties, and has been endorsed by the IVOA Executive Committee as an IVOA Recommendation as of 5 October 2013. It is a stable document and may be used as reference material or cited as a normative reference from another document. IVOA's role in making the Recommendation is to draw attention to the specification and to promote its widespread deployment. This enhances the functionality and interoperability inside the Astronomical Community.

Early versions of this document were known as RegSimpleDAL. The short name is now SimpleDALRegExt.

A list of current IVOA Recommendations and other technical documents can be found at http://www.ivoa.net/Documents/.

## Acknowledgements

This document has been developed with support from the National Science Foundation's Information Technology Research Program under Cooperative Agreement AST0122449 with The Johns Hopkins University, from the UK Particle Physics and Astronomy Research Council (PPARC), and from the Eurpean Commission's Sixth Framework Program via the Optical Infrared Coordination Network (OPTICON).

## Conformance-related definitions

The words "MUST", "SHALL", "SHOULD", "MAY", "RECOMMENDED", and "OPTIONAL" (in upper or lower case) used in this document are to be interpreted as described in IETF standard, RFC 2119 [RFC 2119].

The **Virtual Observatory (VO)** is general term for a collection of federated resources that can be used to conduct astronomical research, education, and outreach. The **International Virtual Observatory Alliance (IVOA)** is a global collaboration of separately funded projects to develop standards and infrastructure that enable VO applications.

XML document **validation** is a software process that checks that an XML document is not only well-formed XML but also conforms to the syntax rules defined by the applicable schema. Typically, when the schema is defined by one or more XML Schema [schema] documents (see next section), validation refers to checking for conformance to the syntax described in those Schema documents. This document describes additional syntax constraints that cannot be enforced solely by the rules of XML Schema; thus, in this document, use of the term validation includes the extra checks that goes beyond common Schema-aware parsers which ensure conformance with this document.

## Syntax Notation Using XML Schema

The Extensible Markup Language, or XML, is document syntax for marking textual information with named tags and is defined by the World Wide Web Consortium (W3C) Recommendation, XML 1.0 [XML]. The set of XML tag names and the syntax rules for their use is referred to as the document schema. One way to formally define a schema for XML documents is using the W3C standard known as XML Schema [schema].

This document defines the VOResource schema using XML Schema. The full Schema document is listed in Appendix A. Parts of the schema appear within the main sections of this document; however, documentation nodes have been left out for the sake of brevity.

Reference to specific elements and types defined in the VOResource schema include the namespaces prefix, `vr`, as in `vr:Resource` (a type defined in the VOResource schema). Reference to specific elements and types defined in the VODataService extension schema include the namespaces prefix, `vs`, as in `vs:ParamHTTP` (a type defined in the VODataService schema). Use of the `vs` prefix in compliant instance documents is strongly recommended, particularly in the applications that involve IVOA Registries (see [RI], section 3.1.2). Elsewhere, the use is not required.

# Contents

# 1. Introduction

Four data access service protocols play a key role in discovering data in the VO:

- **Simple Cone Search** [SCS] -- searches a catalog for sources or observations that are within a given distance of a sky position.
- **Simple Image Access** [SIA] -- searches an archive for images that overlap a given region of sky.
- **Simple Spectral Access** [SSA] -- searches an archive for spectra of positions within a given region of sky.
- **Simple Line Access** [SLA] -- searches a catalog specializing in descriptions of spectral line transitions.

They are called "simple" because a typical query can be formed using only a few search parameters encoded into a URL (i.e. an HTTP GET request). Their power for data discovery comes from the ability of an application to form a single query according to the rules of one of these protocols and send it to multiple services selected, say, for their relevence to a scientific topic which support that protocol. The results collected from those services, in effect then, represent all the relevent data of that type known to the VO. Thus, the key for an application wishing to do a comprehensive search of the VO is to discover all of the services that support the particular standard protocol.

Service discovery in the VO is done via a *searchable registry* [RI]--i.e., a searchable repository of descriptions of resources in VO. These descriptions are comprised of common standard metadata [RM] that capture information about what a resource contains or does and who provides it. A standard registry encodes these descriptions using the VOResource XML Schema [VOR]. Service resources in particular include *capability metadata* that describe the functionality it supports along with *interface metadata* that describe how to access that functionality. It is within the capability metadata that it is possible to indicate support for a particular standard protocol.

Capability metadata play an important role beyond just identifying support for a standard interface. More generally, they describe how the service behaves, and if applications are to make use of this information in an automated way, the behavior must be described using standardized metadata. In general, the metadata necessary for describing that behavior will be specific to the particular kind of service. In the case of a standard protocol, in which it is common that some variation in behavior is allowed while still being in compliance, it can be important to an application to know *how* a service complies with the standard for two reasons:

1. The application may wish to search for and select services that support a particular protocol feature. For example, an application may wish to find image services that can create cut-outs on-the-fly.

2. The application may wish to plan its use of the service according its limitations, such as the maximum region of sky that can be searched in one query.

It is important to note that the relevent behavioral differences between separate services that support a common protocol--and thus the metadata used to describe those behaviors--will be specific to that protocol. That is, for example, the ability to create image cut-outs is irrelevent to the Simple Cone Search protocol. Consequently, it is necessary to define *protocol-specific metadata* to adequately describe a service's support for that protocol. This document defines such capability metadata for SCS, SIA, SSA, and SLA.

This document describes for each of the standard data access protocols--SCS, SIA, SSA, and SLA-- precisely how to describe a service that supports one of the protocols in terms of the VOResource XML encoding standard [VOR]. This specification is intended to be applicable wherever VOResource records are used, but in particular, it is intended as the standard for encoding resource descriptions within an IVOA-compliant registry [RI] and for encoding capability metadata available through the VO Standard Interface [VOSI].

## 1.1. The Role in IVOA Architecture

The IVOA Architecture [Arch] provides a high-level view of how IVOA standards work together to connect users and applications with providers of data and services, as depicted in the diagram in Fig. 1.

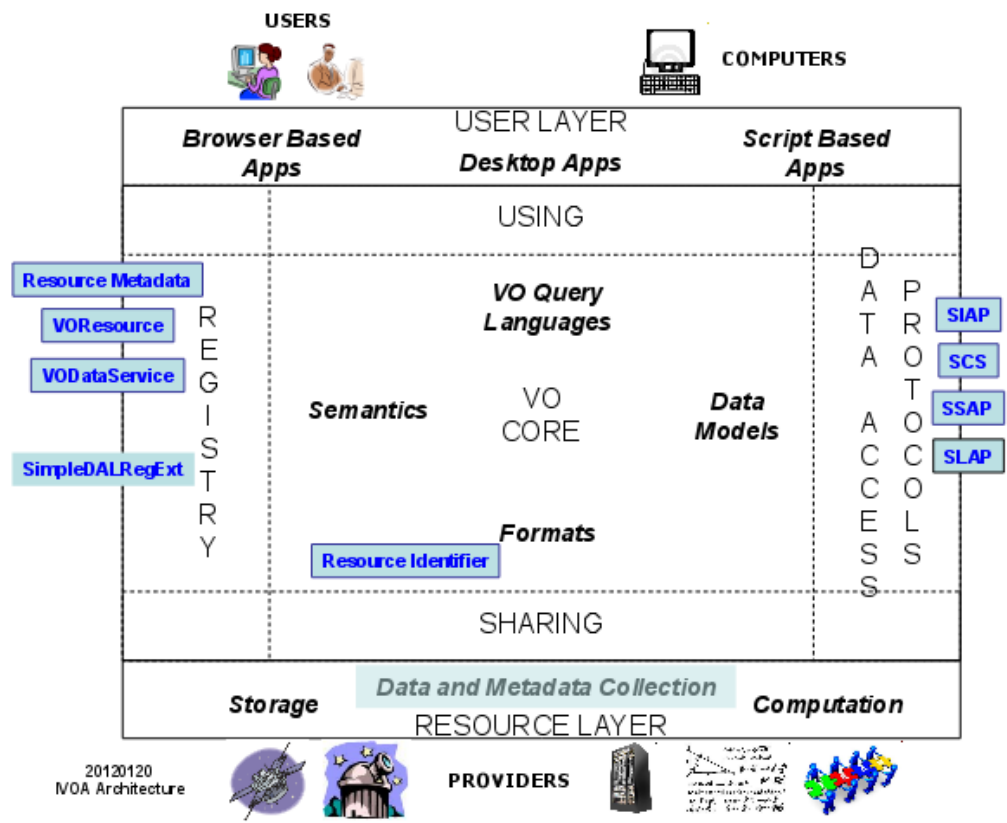

**Figure 1. SimpleDALRegExt in the IVOA Architecture.** The Registry enables applications in the User Layer to discover archives in the Resource Layer and the services they provide for accessing data, particularly those that support the standard data access protocols like SIAP, SCS, SSAP, and SLAP (illustrated on the right). The registry metadata model standards (in blue text and boxes on the left) give structure to the information that enables that discovery. In particular, the SimpleDALRegExt standard defines the metadata used to describe standard data access services of the types listed on the right.

In this architecture, data access protocols provide the means for users (via the User Layer) to access data from archives. Of particular importance are the standard protocols, SCS, SIA, SSA, and SLA, which allow a generic user tool to find data in any archive that supports those protocols. Registries provide to tools in the User Layer a means to discover which archives support the standard protocols. A registry is a repository of descriptions of resources, such as standard services, that can be searched based on the metadata in those descriptions.

Resource descriptions have a well-defined structure: the core concepts are defined in the Resource Metadata standard [RM], and the format is defined by the VOResource XML standard [VOR]. Additional metadata specialized to describe a specific kind of service are defined via extensions to the VOResource core XML Schema. SimpleDALRegExt is one such extension specifically for describing SCS, SIA, SSA, and SLA services in the registry.

## 1.2. Dependencies on Other Standards

This specification relies directly on other IVOA standards in the following ways:

**VOResource, v1.03 [VOR]:**
Descriptions of services that support the standard protocols are encoded using the VOResource XML Schema. The protocol-specific schemas defined in this document are extensions of the VOResource core schema.

**Simple Cone Search, v1.03 [SCS],**
**Simple Image Access, v1.0 [SIA],**
**Simple Spectral Access, v1.04 [SSA],**
**Simple Line Access, v1.0 [SLA]:**
Each protocol specification describes the metadata concepts that should be included in a description of a service that supports the specification.

**VODataService, v1.1 [VDS]:**
The interface to the standard protocol functionality is described with a specialized Interface type, `vs:ParamHTTP`, which is defined in the VODataService XML Schema, an extension to VOResource. This document also recommends describing the service using VODataService resource type, `vs:CatalogDataService`.

Except where noted in subsequent sections, this specification does not imply support for any other versions (including later versions) than the ones noted above.

This specification refers to other IVOA standards:

**Registry Interfaces, v1.0 [RI]:**
A registry that is compliant with both this specification and the Registry Interfaces standard will encode service resource descriptions according to the recommendations in this document.

**IVOA Standard Interface, v1.0 [VOSI]:**
A service that supports one of the target protocols as well as the capability metadata retrieval method defined by VOSI [VOSI, section 2.1] is compliant with this specfcation if the capability metadata are encoded according the recommendations in this document.

Unlike with the previously mentioned specifications, this specification may apply to later versions of the RI and VOSI standards.

# 2. The Common Data Model for Simple DAL Services

This section describes common requirements for describing the target DAL services as VOResource records.

To be recognized as a service, the DAL service resource must be described as a resource type of

`vr:Service` (defined in the VOResource schema [VOR]) or one of its legal sub-types. As specified in the VOResource specification [VOR], the resource type is set by setting the `xsi:type` attribute on the element representing the root of the VOResource record to the namespace-qualified resource type name. As the DAL services respond to queries with tables of available data products, the resource should set the resource type to `vs:CatalogService` (defined in the VODataService extension schema [VDS]). In this case, record authors are encouraged to include a full description of the columns in the table returned in query response (assuming full verbosity). The `vs:CatalogService` resource type also allows the record to provide sky coverage information which authors are also encouraged to provide; an exception to this would be for pure SLA services as the spectral line catalogs they serve are not strictly sky observations.

> **Note:**
> In the future, a more appropriate resource type may be defined for describing DAL services; thus, the loose requirement on the resource type allows for this. At the time of this writing, the `vs:CatalogService` is recommended as the most appropriate type and represents current practice with IVOA registries.

The VOResource record must include a `<capability>` element that describes the services support for the DAL protocol. The contents of the element is described in the next section. In all cases, the value of the `<capability>` element's `standardID` unambiguously identifies the service's support for the particular DAL protocol. The resource may include other `<capability>` elements to describe support for other protocols.

The `<capability>` element describing support for the DAL protocol must include a child `<interface>` element that describes support for the required DAL interface; the `xsi:type` attribute on that element must be set to `vs:ParamHTTP`, and the `role` attribute must be set to "std". A `<accessURL>` element within that `<interface>` must be set to the base URL, as defined in the DAL protocol specification, that provides access to the standard DAL protocol. It is not necessary to provide the `use` attribute to the `<accessURL>` element (as its value can be assumed); however, when it is provided, it must be set to "base". Similarly, it is not necessary to provide the `<interface>` element with `<queryType>` or `<resultType>` elements; however, when provided, their values should be "GET" and "application/x-votable+xml", respectively. The `vs:ParamHTTP` allows one to describe input parameters supported by the service; description authors are encouraged to list the optional parameters and any custom parameters supported by the instance of the service.

---

**A sample interface description for a simple DAL service.**

```
<interface xsi:type="vs:ParamHTTP" role="std">
    <accessURL use="base">
        http://adil.ncsa.uiuc.edu/cgi-bin/voimquery?survey=f&
    </accessURL>

    <!-- here is a standard, optional parameter -->
    <param use="optional" std="true">
      <name>CFRAME</name>
      <description>request to shift to a given coordinate frame.</description>
      <dataType>string</dataType>
    </param>

    <!-- here is a site-specific parameter that this service supports -->
    <param use="optional" std="false">
      <name>FREQ</name>
      <description>Frequency of observation.</description>
      <unit>Hz</unit>
      <dataType>real</dataType>
    </param>
</interface>
```

# 3. Describing Standard Capabilities

This section describes the specific VOResource metadata extension schemas used to describe support for the target DAL protocols. The purpose of these schemas are to provide the `vr:Capability` sub-type that identifies the specific protocol. These are defined employing the recommendations for `vr:Capability` extensions given in the VOResource standard [VR]. In particular, each extension schema has the following features:

- The namespace associated with the extension is a URI that is intended to resolve an HTTP URL to XML Schema document that defines the extension schema. This means that VOResource document authors may use this URI as the location URL within the value of `xsi:schemaLocation` attribute.

> **Note:**
> The IVOA Registry Interface standard [RI] actually *requires* that the VOResource records it shares with other registries provide location URLs via `xsi:schemaLocation` for the VOResource schema and all legal extension schemas that are used in the records. This rule would apply to the extension schemas defined in this standard.

- A particular namespace prefix is recommended for use when referring to the specialized `vr:Capability` sub-type defined in the schema. Generally instance documents may use any prefix; however, in applications where common and consistent use of prefixes is recommended (such as with the Registry Interface specification [RI], use of the prefixes recommended in this document can be used.

- The schema sets `elementFormDefault="unqualified"`. This means that it is not necessary to qualify element names defined in the schema with a namespace prefix (as there are no global elements defined). The only place it is usually needed is as a qualifier to a `Capability` sub-type name given as the value of an `xsi:type` attribute on the `<capability>` element (see the examples in the subsections below).

- The definition of the specialized `vr:Capability` sub-type fixes the value of its `standardID` attribute to the URI that is intended to uniquely identify the standard DAL protocol whose support the type describes.

- The specialized `vr:Capability` sub-type includes a `<testQuery>` element for encoding parameters that together can be used to test the service. The format for encoding the individual parameters is customized for each of the four simple services covered in this specification.

> **Note:**
> It is also possible to encode test queries as part of the `vs:ParamHTTP` interface via its `<testQuery>` element [VDS]; this query is encoded as a single string that can be appended to the service's base URL. This specification does not recommend one over the other. Use of the `vr:Capability`-based `<testQuery>` pre-dates the `vs:ParamHTTP`-based one and has been kept for backward-compatibility purposes. It might be easier for some clients to use since each parameter is individually tagged and perhaps easier to parse, manipulate, and enter automatically into an interface. The `vs:ParamHTTP`-based one has an advantage in that it can be applied to any REST-like service, standard or not.

## 3.1. Simple Cone Search

This section describes the ConeSearch VOResource metadata extension schema which is used to describe services that comply with the Simple Cone Search protocol [SCS].

### 3.1.1. The Schema Namespace

The namespace associated with the ConeSearch extension schema is "http://www.ivoa.net/xml/ConeSearch/v1.0". The namespace prefix, **cs**, should be used in applications where common use of prefixes improves interoperability (e.g. in the IVOA registries [RI]). Furthermore, we use the **cs** prefix in this document to refer to types defined as part of the ConeSearch extension schema.

### 3.1.2. ConeSearch

The **cs:ConeSearch** type is a **vr:Capability** sub-type that should be used to describe a service's support for the Simple Cone Search protocol [SCS]; it is defined as follows:

---

**cs:ConeSearch Type Schema Definition:**
*(Note that the purpose of type CSCapRestriction is to fix the value of standardID.)*

```xml
<xs:complexType name="CSCapRestriction" abstract="true" >
  <xs:complexContent >
    <xs:restriction base="vr:Capability" >
      <xs:sequence >
        <xs:element name="validationLevel" type="vr:Validation"
                    minOccurs="0" maxOccurs="unbounded" />
        <xs:element name="description" type="xs:token" minOccurs="0" />
        <xs:element name="interface" type="vr:Interface"
                    minOccurs="0" maxOccurs="unbounded" />
      </sequence>
      <xs:attribute name="standardID" type="vr:IdentifierURI" use="required"
                    fixed="ivo://ivoa.net/std/ConeSearch" />
    </restriction>
  </complexContent>
</complexType>

<xs:complexType name="ConeSearch" >
  <xs:complexContent >
    <xs:extension base="cs:CSCapRestriction" >
      <xs:sequence >
        <xs:element name="maxSR" type="xs:float" minOccurs="0" maxOccurs="1"/>
        <xs:element name="maxRecords" type="xs:positiveInteger"
                    minOccurs="0" maxOccurs="1"/>
        <xs:element name="verbosity" type="xs:boolean" />
        <xs:element name="testQuery" type="cs:Query" minOccurs="0" maxOccurs="1"/>
      </sequence>
    </extension>
  </complexContent>
</complexType>
```

---

The **cs:CSCapRestriction** is defined to fix the value of the **standardID** attribute; thus, all uses of the **cs:ConeSearch** type must set **standardID** to "ivo://ivoa.net/std/ConeSearch". Because the **cs:CSCapRestriction** is marked as abstract, instance documents can not use it directly as a value for the **xsi:type** attribute.

The custom metadata that the **cs:ConeSearch** type provides is given in the table below. For the elements whose semantics map directly to service profile metadata called for in the SCS standard [SCS, section 3], there is an entry labeled "SCS Name"; this indicates the metadata name given in the SCS specification that the element in this schema corresponds to. The profile metadata listed in the SCS specification that is not covered by the elements below are covered by other metadata that are part of the core VOResource schema.

| cs:ConeSearch Extension Metadata Elements | |
|---|---|
| **Element** | **Definition** |
| maxSR | *SCS Name:*      MaxSR |
| | *Value type:*      floating-point number: `xs:float` |
| | *Semantic Meaning:* The largest search radius, in degrees, that will be accepted by the service without returning an error condition. Not providing this element or specifying a value of 180 indicates that there is no restriction. |
| | *Occurrences:*      optional |
| | *Comments:*      Not providing a value is the prefered way to indicate that there is no restriction. |
| maxRecords | *SCS Name:*      MaxRecords |
| | *Value type:*      positive integer: `xs:positiveInteger` |
| | *Semantic Meaning:* The largest number of records that the service will return. Not providing this value means that there is no effective limit. |
| | *Occurrences:*      optional |
| | *Comments:*      This does not refer to the total number of records in the catalog but rather maximum number of records the service is capable of returning. A limit that is greater than the number of records available in the archive is equivalent to their being no effective limit. (See RM, Hanisch 2007.) |
| verbosity | *SCS Name:*      Verbosity |
| | *Value type:*      boolean (true/false): `xs:boolean` |
| | *Semantic Meaning:* True if the service supports the VERB keyword; false, otherwise. |
| | *Occurrences:*      required |
| | *Comments:*      The value should be false if the all values of the VERB input parameter results in the same set of columns being returned. |
| testQuery | *Value type:*      composite: `cs:Query` |
| | *Semantic Meaning:* A query that will result in at least one matched record that can be used to test the service. |
| | *Occurrences:*      optional |

### 3.1.2.1 testQuery and the Query Type

The `<testQuery>` element is intended to help other VO components (e.g. registries, validation services, services that monitor the VO's operational health--but typically not end users) test that the service is up and operating correctly. It provides a set of legal input parameters that should return a legal response that includes at least one matched record. Since this query is intended for testing purposes, the size of the result set should be small.

The `cs:Query` type captures the different components of the query into separate elements, as defined below:

**cs:Query Type Schema Definition**

```
<xs:complexType name="Query" >
  <xs:sequence >
    <xs:element name="ra" type="xs:double" />
    <xs:element name="dec" type="xs:double" />
    <xs:element name="sr" type="xs:double" />
    <xs:element name="verb" type="xs:positiveInteger" minOccurs="0" />
    <xs:element name="catalog" type="xs:string" minOccurs="0" />
    <xs:element name="extras" type="xs:string" minOccurs="0" />
  </xs:sequence>
</xs:complexType>
```

The individual sub-elements are defined as follows:

**cs:Query Metadata Elements**

| Element | Definition | |
|---|---|---|
| ra | *Value type:* | floating-point number: `xs:double` |
| | *Semantic Meaning:* | the right ascension of the search cone's center in decimal degrees. |
| | *Occurrences:* | required |
| dec | *Value type:* | floating-point number: `xs:double` |
| | *Semantic Meaning:* | the declination of the search cone's center in decimal degrees. |
| | *Occurrences:* | required |
| sr | *Value type:* | floating-point number: `xs:double` |
| | *Semantic Meaning:* | the radius of the search cone in decimal degrees. |
| | *Occurrences:* | required |
| verb | *Value type:* | positive integer: `xs:positiveInteger` |
| | *Semantic Meaning:* | the verbosity level to use where 1 or less means the bare minimum set of columns and 3 or more means the full set of available columns. |
| | *Occurrences:* | optional |
| catalog | *Value type:* | string: `xs:string` |
| | *Semantic Meaning:* | the catalog to query. |
| | *Occurrences:* | optional |
| | *Comments:* | When the service can access more than one catalog, this input parameter, if available, is used to indicate which service to access. |
| extras | *Value type:* | string: `xs:string` |
| | *Semantic Meaning:* | any extra (non-standard) parameters that must be provided (apart from what is part of base URL given by the accessURL element). |
| | *Occurrences:* | optional |
| | *Comments:* | this value should be in the form of name=value pairs delimited with apersands (&). |

## 3.2. Simple Image Access

This section describes the SIA VOResource metadata extension schema which is used to describe services that comply with the Simple Image Access protocol [SIA].

### 3.2.1. The Schema Namespace

The namespace associated with the SIA extension schema is "http://www.ivoa.net/xml/SIA/v1.1". The namespace prefix, `sia`, should be used in applications where common use of prefixes improves interoperability (e.g. in the IVOA registries [RI]). Furthermore, we use the `sia` prefix in this document to refer to types defined as part of the SIA extension schema.

### 3.2.2. SimpleImageAccess

The `sia:SimpleImageAccess` type is a `vr:Capability` sub-type that should be used to describe a service's support for the Simple Image Access protocol [SIA]; it is defined as follows:

---

**sia:SimpleImageAccess Type Schema Definition**
*(Note that the purpose of type `SIACapRestriction` is to fix the value of `standardID`.)*

```
<xs:complexType name="SIACapRestriction" abstract="true" >
  <xs:complexContent >
    <xs:restriction base="vr:Capability" >
      <xs:sequence >
        <xs:element name="validationLevel" type="vr:Validation"
                    minOccurs="0" maxOccurs="unbounded" />
        <xs:element name="description" type="xs:token" minOccurs="0" />
        <xs:element name="interface" type="vr:Interface"
                    minOccurs="0" maxOccurs="unbounded" />
      </xs:sequence>
      <xs:attribute name="standardID" type="vr:IdentifierURI" use="required"
                    fixed="ivo://ivoa.net/std/SIA" />
    </xs:restriction>
  </xs:complexContent>
</xs:complexType>

<xs:complexType name="SimpleImageAccess" >
  <xs:complexContent >
    <xs:extension base="sia:SIACapRestriction" >
      <xs:sequence >
        <xs:element name="imageServiceType" type="sia:ImageServiceType" />
        <xs:element name="maxQueryRegionSize" type="sia:SkySize" minOccurs="0" maxOccurs="1"/>
        <xs:element name="maxImageExtent" type="sia:SkySize" minOccurs="0" maxOccurs="1"/>
        <xs:element name="maxImageSize" type="xs:positiveInteger" minOccurs="0" maxOccurs="1"/>
        <xs:element name="maxFileSize" type="xs:positiveInteger" minOccurs="0" maxOccurs="1"/>
        <xs:element name="maxRecords" type="xs:positiveInteger" minOccurs="0" maxOccurs="1"/>
        <xs:element name="testQuery" type="sia:Query" minOccurs="0" maxOccurs="1" />
      </xs:sequence>
    </xs:extension>
  </xs:complexContent>
</xs:complexType>
```

---

The `sia:SIACapRestriction` is defined to fix the value of the `standardID` attribute; thus, all uses of the `sia:SimpleImageAccess` type must set `standardID` to "ivo://ivoa.net/std/SIA". Because the `sia:SIACapRestriction` is marked as abstract, instance documents can not use it directly as a value for the `xsi:type` attribute.

The custom metadata that the `sia:SimpleImageAccess` type provides is given in the table below. For the elements whose semantics map directly to metadata called for in the SIA standard [SIA, section 7], there is an entry labeled "SIA Name"; this indicates the metadata name given in the SIA specification that the element in this schema corresponds to.

| sia:SimpleImageAccess Extension Metadata Elements | |
|---|---|
| **Element** | **Definition** |
| imageServiceType | *SIA Name:* Type.ImageService |
| | *Value type:* string with controlled vocabulary |
| | *Semantic Meaning:* The category of image service as defined by [SIA], section 3. |
| | *Occurrences:* required |
| | *Allowed Values:* `Cutout` an Image Cutout Service, as defined in [SIA], section 3 (see Note below). |
| | `Mosaic` an Image Mosaicing Service, as defined in [SIA], section 3 (see Note below). |
| | `Atlas` an Atlas Image Archive, as defined in [SIA], section 3 (see Note below). |
| | `Pointed` a Pointed Image Archive, as defined in [SIA], section 3 (see Note below). |
| maxQueryRegionSize | *SIA Name:* MaxQueryRegionSize |
| | *Value type:* composite: `sia:SkySize` |
| | *Semantic Meaning:* The maximum image query region size, expressed in decimal degrees. Not providing this element or specifying a value of 360 degrees indicates that there is no limit and the entire data collection (entire sky) can be queried. |
| | *Occurrences:* optional |
| | *Comments:* Not providing a value is the prefered way to indicate that there is no limit. |
| maxImageExtent | *SIA Name:* MaxImageExtent |
| | *Value type:* composite: `sia:SkySize` |
| | *Semantic Meaning:* An upper bound on a region of the sky that can be covered by returned images. That is, no image returned by this service will cover more than this limit. Not providing this element or specifying a value of 360 degrees indicates that there is no fundamental limit to the region covered by a returned image. |
| | *Occurrences:* optional |
| | *Comments:* When the `<imageServiceType>` is "Cutout" or "Mosaic", this represents the largest area that can be requested. In this case, the "no limit" value means that all-sky images can be requested. When the type is "Atlas" or "Pointed", it should be a region that most closely encloses largest images in the archive, and the "no limit" value means that the archive contains all-sky (or nearly so) images.<br><br>Not providing a value is the prefered way to indicate that there is no limit. |

*continued next page*

| sia:SimpleImageAccess Extension Metadata Elem ents (con't) | |
|---|---|
| **Element** | **Definition** |
| maxImageSize | *SIA Name:*    MaxImageSize<br><br>*Value type:*    positive integer: `xs:positiveInteger`<br><br>*Semantic Meaning:* A measure of the largest image the service can produce given as the maximum number of pixels along the first or second axes. Not providing a value indicates that there is no effective limit to the size of the images that can be returned.<br><br>*Occurrences:*    optional<br><br>*Comments:*    This is primarily relevant when the <imageServiceType> is "Cutout" or "Mosaic", indicating the largest image that can be created. When the <imageServiceType> is "Atlas" or "Pointed", this should be specified only when there are static images in the archive that can be searched for but not returned because they are too big.<br><br>When a service is more fundamentally limited by the total number of pixels in the image, this value should be set to the square-root of that number. This number will then represent a lower limit on the maximum length of a side. |
| maxFileSize | *SIA Name:*    MaxFileSize<br><br>*Value type:*    positive integer: `xs:positiveInteger`<br><br>*Semantic Meaning:* The maximum image file size in bytes. Not providing a value indicates that there is no effective limit the size of files that can be returned.<br><br>*Occurrences:*    optional<br><br>*Comments:*    This is primarily relevant when the <imageServiceType> is "Cutout" or "Mosaic", indicating the largest files that can be created. When the <imageServiceType> is "Atlas" or "Pointed", this should be specified only when there are static images in the archive that can be searched for but not returned because they are too big. |
| maxRecords | *SIA Name:*    MaxRecords<br><br>*Value type:*    positive integer: `xs:positiveInteger`<br><br>*Semantic Meaning:* The largest number of records that the Image Query web method will return. Not providing this value means that there is no effective limit.<br><br>*Occurrences:*    optional<br><br>*Comments:*    This does not refer to the total number of images in the archive but rather maximum number of records the service is capable of returning. A limit that is greater than the number of images available in the archive is equivalent to their being no effective limit. (See also [RM], Sect. 5.2.) |
| testQuery | *Value type:*    composite: `sia:Query`<br><br>*Semantic Meaning:* a set of query parameters that is expected to produce at least one matched record which can be used to test the service.<br><br>*Occurrences:*    optional |

The `sia:ImageServiceType` type is provided to restrict the values of the <imageServiceType> element to those allowed by the SIA standard:

**sia:ImageServiceType Type Schema Definition**

```
<xs:simpleType name="ImageServiceType" >
  <xs:restriction base="xs:token" >
    <xs:enumeration value="Cutout" />
    <xs:enumeration value="Mosaic" />
    <xs:enumeration value="Atlas" />
    <xs:enumeration value="Pointed" />
  </xs:restriction>
</simpleType>
```

**Note:**

The SIA specification defines the image service types as follows:

***Image Cutout Service***

This is a service which extracts or "cuts out" rectangular regions of some larger image, returning an image of the requested size to the client. Such images are usually drawn from a database or a collection of survey images that cover some large portion of the sky. To be considered a cutout service, the returned image should closely approximate (or at least not exceed) the size of the requested region; however, a cutout service will not normally resample (rescale or reproject) the pixel data. A cutout service may mosaic image segments to cover a large region but is still considered a cutout service if it does not resample the data. Image cutout services are fast and avoid image degredation due to resampling.

***Image Mosaicing Service***

This service is similar to the image cutout service but adds the capability to compute an image of the size, scale, and projection specified by the client. Mosaic services include services which resample and reproject existing image data, as well as services which generate pixels from some more fundamental dataset, e.g., a high energy event list or a radio astronomy measurement set. Image mosaics can be expensive to generate for large regions but they make it easier for the client to overlay image data from different sources. Image mosaicing services which resample already pixelated data will degrade the data slightly, unlike the simpler cutout service which returns the data unchanged.

***Atlas Image Archive***

This category of service provides access to pre-computed images that make up a survey of some large portion of the sky. The service, however, is not capable of dynamically cutting out requested regions, and the size of atlas images is predetermined by the survey. Atlas images may range in size from small cutouts of extended objects to large calibrated survey data frames.

***Pointed Image Archive***

This category of service provides access to collections of images of many small, "pointed" regions of the sky. "Pointed" images normally focus on specific sources in the sky as opposed to being part of a sky survey. This type of service usually applies to instrumental archives from observatories with guest observer programs (e.g., the HST archive) and other general purpose image archives (e.g., the ADIL). If a service provides access to both survey and pointed images, then it should be considered a Pointed Image Archive for the purposes of this specification; if a differentiation between the types of data is desired the pointed and survey data collections should be registered as separate image services.

Several of the `sia:SimpleImageAccess` metadata use complex types to capture their values; the subsequent sections below define those special types.

### 3.2.2.1. SkySize

The `sia:SkySize` type is used to capture simple rectangular extents on the sky along longitudinal and latitudinal directions. It is defined as follows:

---
**sia:SkySize Type Schema Definition**

```
<xs:complexType name="SkySize" >
  <xs:sequence >
    <xs:element name="long" type="xs:double" />
    <xs:element name="lat" type="xs:double" />
  </sequence>
</complexType>
```
---

The coordinate system for the region is intended to be implied by the context of its use--i.e. the definition of the element defined with this type. In this SIA case, the longitudinal and latitudinal values represent the extents along right ascension and declination in the ICRS system (the system assumed by the SIA interface).

---
**sia:SkySize Metadata Elements**

| Element | Definition | |
|---------|------------|--|
| long | *Value type:* | floating-point number: `xs:double` |
| | *Semantic Meaning:* | The maximum size in the longitude (R.A.) direction given in degrees |
| | *Occurrences:* | required |
| lat | *Value type:* | floating-point number: `xs:double` |
| | *Semantic Meaning:* | The maximum size in the latitude (Dec.) direction given in degrees |
| | *Occurrences:* | required |
---

### 3.2.2.3. testQuery and the Query Type

As with the other DAL `vr:capability` types, the `<testQuery>` element is intended to help other VO components (e.g. registries, validation services, services that monitor the VO's operational health--but typically not end users) test that the service is up and operating correctly. It provides a set of legal input parameters that should return a legal response that includes at least matched record. Since this query is intended for testing purposes, the size of the result set should be small.

The `sia:Query` type captures the different components of the query into separate elements, as defined below:

---
**sia:Query Type Schema Definition**

```
<xs:complexType name="Query" >
  <xs:sequence >
    <xs:element name="pos" type="sia:SkyPos" minOccurs="0" />
    <xs:element name="size" type="sia:SkySize" minOccurs="0" />
    <xs:element name="verb" type="xs:positiveInteger" minOccurs="0" />
    <xs:element name="extras" type="xs:string" minOccurs="0" />
  </sequence>
</complexType>
```
---

The individual sub-elements are defined as follows:

| sia:Query Metadata Elements | | |
| --- | --- | --- |
| **Element** | **Definition** | |
| pos | *Value type:* | composite: sia:SkyPos |
| | *Semantic Meaning:* | the center position of the rectangular region that should be used as part of the query to the SIA service. |
| | *Occurrences:* | optional |
| size | *Value type:* | composite: sia:SkySize |
| | *Semantic Meaning:* | the rectangular size of the region that should be used as part of the query to the SIA service. |
| | *Occurrences:* | optional |
| verb | *Value type:* | positive integer: `xs:positiveInteger` |
| | *Semantic Meaning:* | the verbosity level to use where 1 or less means the bare minimum set of columns and 3 or more means the full set of available columns. |
| | *Occurrences:* | optional |
| extras | *Value type:* | string: `xs:string` |
| | *Semantic Meaning:* | any extra (particularly non-standard) parameters that must be provided (apart from what is part of base URL given by the accessURL element). |
| | *Occurrences:* | optional |
| | *Comments:* | this value should be in the form of name=value pairs delimited with apersands (&, properly escaped for inclusion into XML). |

### 3.2.2.4. SkyPos

The `sia:SkyPos` type is used to encode the `<testQuery>`'s `<pos>` element, the center position of the test region of interest.

| sia:SkyPos Type Schema Definition |
| --- |
| ```xml
<xs:complexType name="SkyPos" >
  <xs:sequence >
    <xs:element name="long" type="xs:double" />
    <xs:element name="lat" type="xs:double" />
  </xs:sequence>
</xs:complexType>
``` |

| sia:SkyPos Metadata Elements | | |
| --- | --- | --- |
| **Element** | **Definition** | |
| long | *Value type:* | floating-point number: `xs:double` |
| | *Semantic Meaning:* | The sky position in the longitude (R.A.) direction |
| | *Occurrences:* | required |
| lat | *Value type:* | floating-point number: `xs:double` |
| | *Semantic Meaning:* | The sky position in the latitude (Dec.) direction |
| | *Occurrences:* | required |

## 3.3. Simple Spectral Access

This section describes the SSA VOResource metadata extension schema which is used to describe services that comply with the Simple Spectral Access protocol. This differs from the other Simple DAL extensions in that it defines two `vr:Capability` types: `ssap:SimpleSpectralAccess` and `ssap:ProtoSpectralAccess`. The former is intended for services that are intended to be compliant with published SSA Recommendation [SSA]. The latter is intended for services that were deployed prior to the publication of the SSA Recommendation (see section 3.3.3, below).

### 3.3.1. The Schema Namespace

The namespace associated with the SSA extension schema is "http://www.ivoa.net/xml/SSA/v1.1". The namespace prefix, `ssap`, should be used in applications where common use of prefixes improves interoperability (e.g. in the IVOA registries [RI]). Furthermore, we use the `ssap` prefix in this document to refer to types defined as part of the SSA extension schema.

> **Note:**
> Though it departs a bit from convention, the `ssap` prefix was chosen to avoid a collision with its use in [SSA] for identifying UTypes from the Spectral Data Model.

### 3.3.2. SimpleSpectralAccess

The `ssap:SimpleSpectralAccess` type is the `vr:Capability` sub-type that should be used to describe a service's support for the Simple Spectral Access protocol [SSA]; it is defined as follows:

---

**ssap:SimpleSpectralAccess Type Schema Definition**

```xml
<xs:complexType name="SSACapRestriction" abstract="true" >
  <xs:complexContent >
    <xs:restriction base="vr:Capability" >
      <xs:sequence >
        <xs:element name="validationLevel" type="vr:Validation" minOccurs="0"
            maxOccurs="unbounded" />
        <xs:element name="description" type="xs:token" minOccurs="0" />
        <xs:element name="interface" type="vr:Interface" minOccurs="0"
            maxOccurs="unbounded" />
      </xs:sequence>
      <xs:attribute name="standardID" type="vr:IdentifierURI" use="required"
                    fixed="ivo://ivoa.net/std/SSA" />
    </xs:restriction>
  </xs:complexContent>
</xs:complexType>
```

---

*continued on next page*

**ssap:SimpleSpectralAccess Type Schema Definition (con't)**

```
<xs:complexType name="SimpleSpectralAccess" >
  <xs:complexContent >
    <xs:extension base="ssap:SSACapRestriction" >
      <xs:sequence >
        <xs:element name="complianceLevel" type="ssap:ComplianceLevel" />
        <xs:element name="dataSource" type="ssap:DataSource"
                    minOccurs="1" maxOccurs="unbounded" />
        <xs:element name="creationType" type="ssap:CreationType"
                    minOccurs="1" maxOccurs="unbounded" />
        <xs:element name="supportedFrame" type="ssap:SupportedFrame"
                    minOccurs="1" maxOccurs="unbounded"/>
        <xs:element name="maxSearchRadius" type="xs:double" minOccurs="0" maxOccurs="1" />
        <xs:element name="maxRecords" type="xs:positiveInteger" minOccurs="0" maxOccurs="1" />
        <xs:element name="defaultMaxRecords" type="xs:positiveInteger"
                    minOccurs="0" maxOccurs="1" />
        <xs:element name="maxAperture" type="xs:double" minOccurs="0" maxOccurs="1" />
        <xs:element name="maxFileSize" type="xs:positiveInteger" minOccurs="0" maxOccurs="1" />
        <xs:element name="testQuery" type="ssap:Query" minOccurs="0" maxOccurs="1" />
      </xs:sequence>
    </xs:extension>
  </xs:complexContent>
</xs:complexType>
```

The `ssap:SSACapRestriction` is defined to fix the value of the `standardID` attribute; thus, all uses of the `ssap:SimpleSpectralAccess` type must set `standardID` to "ivo://ivoa.net/std/SSA". Because the `cs:SSACapRestriction` is marked as abstract, instance documents can not use it directly as a value for the `xsi:type` attribute.

The custom metadata that the `ssap:SimpleSpectralAccess` type provides is given in the table below. Note that some of these elements derive from the SSA standard [SSA]; others, from the RM standard [RM]. The "Semantic Meaning" entry provides the reference to the original definition.

**ssap:SimpleSpectralAccess Extension Metadata Elements**

| Element | Definition |
|---------|------------|
| complianceLevel | *Value type:*     string with controlled vocabulary<br><br>*Semantic Meaning:* The category indicating the level to which this instance complies with the SSA standard, as defined in [SSA], section 1.4.1.<br><br>*Occurrences:*     required<br><br>*Allowed Values:*    **query**    The service supports all of the capabilities and features of the SSA protocol identified as "must" in the specification, except that it does not support returning data in at least one SSA-compliant format.<br><br>                **minimal**   The service supports all of the capabilities and features of the SSA protocol identified as "must" in the specification.<br><br>                **full**      The service supports all of the capabilities and features of the SSA protocol identified as "must" or "should" in the specification. |

*continued next page*

| ssap:SimpleSpectralAccess Extension Metadata Elements (con't) | |
|---|---|
| **Element** | **Definition** |
| dataSource | *Value type:* string with controlled vocabulary |
| | *Semantic Meaning:* The category specifying where the data originally came from, as defined in [SSA], section 2.5.1. |
| | *Occurrences:* required; multiple occurrences allowed. |
| | *Allowed Values:* |
| | `survey` A survey dataset, which typically covers some region of observational parameter space in a uniform fashion, with as complete as possible coverage in the region of parameter space observed. |
| | `pointed` A pointed observation of a particular astronomical object or field. |
| | `custom` Data which has been custom processed, e.g., as part of a specific research project. |
| | `theory` Theory data, or any data generated from a theoretical model, for example a synthetic spectrum. |
| | `artificial` Artificial or simulated data. |



| ssap:SimpleSpectralAccess Extension Metadata Elements (con't) | |
|---|---|
| **Element** | **Definition** |
| creationType | *Value type:*      string with controlled vocabulary |

| | | |
|---|---|---|
| | *Value type:* | string with controlled vocabulary |
| | *Semantic Meaning:* | The category that describes the process used to produce the dataset, as defined in [SSA], section 2.5.2. |
| | *Occurrences:* | required; multiple occurrences allowed. |
| | *Allowed Values:* | `archival`    The entire archival or project dataset is returned. Transformations such as metadata or data model mediation or format conversions may take place, but the content of the dataset is not substantially modified (e.g., all the data is returned and the sample values are not modified). |
| | | `cutout`    The dataset is subsetted in some region of parameter space to produce a subset dataset. Sample values are not modified, e.g., cutouts could be recombined to reconstitute the original dataset. |
| | | `filtered`    The data is filtered in some fashion to exclude portions of the dataset, e.g., passing only data in selected regions along a measurement axis, or processing the data in a way which recomputes the sample values, e.g., due to interpolation or flux transformation. |
| | | `mosaic`    Data from multiple non- or partially-overlapping datasets are combined to produce a new dataset. |
| | | `projection`    Data is geometrically warped or dimensionally reduced by projecting through a multidimensional dataset. |
| | | `spectralExtraction`    Extraction of a spectrum from another dataset, e.g., extraction of a spectrum from a spectral data cube through a simulated aperture. |
| | | `catalogExtraction`    Extraction of a catalog of some form from another dataset, e.g., extraction of a source catalog from an image, or extraction of a line list catalog from a spectrum (not valid for a SSA service). |
| | *Comments:* | Typically this describes only the processing performed by the data service, but it could describe some additional earlier processing as well, e.g., if data is partially precomputed. |



| ssap:SimpleSpectralAccess Extension Metadata Elements (con't) | |
|---|---|
| **Element** | **Definition** |
| supportedFrame | *Value type:* string with controlled vocabulary |
| | *Semantic Meaning:* The [STC] name of the space-time coordinate frames that this service can accept in. Values given here can be used as the coordinate frame argument to the SSA POS query parameter (See [SSA], secion 4.1.1.1). |
| | *Occurrences:* required; multiple occurrences allowed. |
| | *Allowed Values:* **FK4, FK5, ECLIPTIC, ICRS, GALACTIC_I, GALACTIC_II, SUPER_GALACTIC, AZ_EL, BODY, GEO_C, GEO_D, MAG, GSE, GSM, SM, HGC, HGS, HEEQ, HRTN, HPC, HPR, HCC, HGI, MERCURY_C, VENUS_C, LUNA_C, MARS_C, JUPITER_C_III, SATURN_C_III, URANUS_C_III, NEPTUNE_C_III, PLUTO_C, MERCURY_G, VENUS_G, LUNA_G, MARS_G, JUPITER_G_III, SATURN_G_III, URANUS_G_III, NEPTUNE_G_III, PLUTO_G, UNKNOWN** ; their meanings are defined in Table 3 (section 4.4.1.2.3) of [STC]. |
| | *Comments:* Since the [SSA] standard requires that ICRS be supported by default, the list of supported frame provided here must include at least ICRS. |
| maxSearchRadius | *Value type:* floating-point number: `xs:double` |
| | *Semantic Meaning:* The largest search radius, in degrees, that will be accepted by the service without returning an error condition (as defined in [RM], section 5.2). Not providing this element or specifying a value of 180 indicates that there is no restriction. |
| | *Occurrences:* optional |
| | *Comments:* Not providing a value is the prefered way to indicate that there is no restriction. |
| maxRecords | *Value type:* positive integer: `xs:positiveInteger` |
| | *Semantic Meaning:* The hard limit on the largest number of records that the query operation will return in a single response (as defined in [RM], section 5.2). Not providing this value means that there is no effective limit. |
| | *Occurrences:* optional |
| | *Comments:* This does not refer to the total number of images in the archive but rather maximum number of records the service is capable of returning. A limit that is greater than the number of images available in the archive is equivalent to their being no effective limit. (See also [RM], Sect. 5.2.) |
| defaultMaxRecords | *Value type:* |
| | *Semantic Meaning:* The largest number of records that the service will return when the MAXREC parameter not specified in the query input. Not providing a value means that the hard limit implied by `<maxRecords>` will be the default limit. |
| | *Occurrences:* optional |



| ssap:SimpleSpectralAccess Extension Metadata Elements (con't) | |
|---|---|
| **Element** | **Definition** |
| maxAperture | *Value type:* floating-point number: `xs:double` |
| | *Semantic Meaning:* The largest aperture diameter that can be supported upon request via the APERTURE input parameter by a service that supports the special extraction creation method. A value of 360 or not providing a value means there is no theoretical limit. |
| | *Occurrences:* optional |
| | *Comments:* Not providing a value is the preferred way to indicate that there is no limit. |
| maxFileSize | *Value type:* positive integer: `xs:positiveInteger` |
| | *Semantic Meaning:* The maximum image file size in bytes. |
| | *Occurrences:* optional |
| testQuery | *Value type:* composite: ssap:Query |
| | *Semantic Meaning:* a set of query parameters that is expected to produce at least one matched record which can be used to test the service. |
| | *Occurrences:* optional |

The controlled vocabulary given for the above elements are set by their respective simple types:

| ssap:ComplianceLevel Type Schema Definition |
|---|

```
<xs:simpleType name="ComplianceLevel" >
  <xs:restriction base="xs:token" >
    <xs:enumeration value="query" />
    <xs:enumeration value="minimal" />
    <xs:enumeration value="full" />
  </xs:restriction>
</simpleType>
```

| ssap:DataSource Type Schema Definition |
|---|

```
<xs:simpleType name="DataSource" >
  <xs:restriction base="xs:token" >
    <xs:enumeration value="survey" />
    <xs:enumeration value="pointed" />
    <xs:enumeration value="custom" />
    <xs:enumeration value="theory" />
    <xs:enumeration value="artificial" />
  </xs:restriction>
</simpleType>
```

**ssap:CreationType Type Schema Definition**

```
<xs:simpleType name="CreationType" >
  <xs:restriction base="xs:token" >
    <xs:enumeration value="archival" />
    <xs:enumeration value="cutout" />
    <xs:enumeration value="filtered" />
    <xs:enumeration value="mosaic" />
    <xs:enumeration value="projection" />
    <xs:enumeration value="spectralExtraction" />
    <xs:enumeration value="catalogExtraction" />
  </restriction>
</simpleType>
```

**ssap:SupportedFrame Type Schema Definition**

```
<xs:simpleType name="SupportedFrame" >
  <xs:restriction base="xs:token" >
        <xs:enumeration value="FK4"/>
        <xs:enumeration value="FK5"/>
        <xs:enumeration value="ECLIPTIC"/>
        <xs:enumeration value="ICRS"/>
        <xs:enumeration value="GALACTIC_I"/>
        <xs:enumeration value="GALACTIC_II"/>
        <xs:enumeration value="SUPER_GALACTIC"/>
        <xs:enumeration value="AZ_EL"/>
        <xs:enumeration value="BODY"/>
        <xs:enumeration value="GEO_C"/>
        <xs:enumeration value="GEO_D"/>
        <xs:enumeration value="MAG"/>
        <xs:enumeration value="GSE"/>
        <xs:enumeration value="GSM"/>
        <xs:enumeration value="HGC"/>
        <xs:enumeration value="HGS"/>
        <xs:enumeration value="HEEQ"/>
        <xs:enumeration value="HRTN"/>
        <xs:enumeration value="HPC"/>
        <xs:enumeration value="HPR"/>
        <xs:enumeration value="HCC"/>
        <xs:enumeration value="HGI"/>
        <xs:enumeration value="MERCURY_C"/>
        <xs:enumeration value="VENUS_C"/>
        <xs:enumeration value="LUNA_C"/>
        <xs:enumeration value="MARS_C"/>
        <xs:enumeration value="JUPITER_C_III"/>
        <xs:enumeration value="SATURN_C_III"/>
        <xs:enumeration value="URANUS_C_III"/>
        <xs:enumeration value="NEPTUNE_C_III"/>
        <xs:enumeration value="PLUTO_C"/>
        <xs:enumeration value="MERCURY_G"/>
        <xs:enumeration value="VENUS_G"/>
        <xs:enumeration value="LUNA_G"/>
        <xs:enumeration value="MARS_G"/>
        <xs:enumeration value="JUPITER_G_III"/>
        <xs:enumeration value="SATURN_G_III"/>
        <xs:enumeration value="URANUS_G_III"/>
        <xs:enumeration value="NEPTUNE_G_III"/>
        <xs:enumeration value="PLUTO_G"/>
        <xs:enumeration value="UNKNOWN"/>
  </restriction>
</simpleType>
```

The definitions of the coordinate reference frames identified in the **ssap:SupportedFrame** type are the same given in [STC] (and the references therein).

### 3.3.2.1 testQuery and the Query Type

As with the other DAL `vr:capability` types, the `<testQuery>` element is intended to help other VO components (e.g. registries, validation services, services that monitor the VO's operational health--but typically not end users) test that the service is up and operating correctly. It provides a set of legal input parameters that should return a legal response that includes at least matched record. Since this query is intended for testing purposes, the size of the result set should be small.

The `ssap:Query` type captures the different components of the query into separate elements, as defined below:

| ssap:Query Type Schema Definition |
|---|
| ```
<xs:complexType name="Query" >
  <xs:sequence >
    <xs:element name="pos" type="ssap:PosParam" minOccurs="0" />
    <xs:element name="size" type="xs:double" minOccurs="0" />
    <xs:element name="queryDataCmd" type="xs:string" minOccurs="0" />
  </xs:sequence>
</xs:complexType>
``` |

The individual sub-elements are defined as follows:

| ssap:Query Metadata Elements | |
|---|---|
| **Element** | **Definition** |
| pos | *Value type:* composite: `ssap:PosParam` <br> *Semantic Meaning:* the center position the search cone given in decimal degrees. <br> *Occurrences:* optional |
| size | *Value type:* floating-point number: `xs:double` <br> *Semantic Meaning:* the size of the search radius. <br> *Occurrences:* optional |
| queryDataCmd | *Value type:* string: `xs:string` <br> *Semantic Meaning:* Fully specified test query formatted as an URL argument list in the syntax specified by the SSA standard. The list must exclude the REQUEST argument which is assumed to be set to "queryData". <br> *Occurrences:* optional <br> *Comments:* This value must be in the form of name=value pairs delimited with apersands (&). A query may then be formed by appending to the base URL the request argument, "REQUEST=queryData&", followed by the contents of this element. |

### 3.3.2.2. PosParam

The `ssap:PosParam` type is used to encode the `<testQuery>`'s `<pos>` element, the center position of the test region of interest; it is defined as follows:

| ssap:PosParam Type Schema Definition |
|---|
| ```
<xs:complexType name="PosParam" >
  <xs:sequence >
    <xs:element name="long" type="xs:double" />
    <xs:element name="lat" type="xs:double" />
    <xs:element name="refframe" type="xs:token" minOccurs="0" />
  </xs:sequence>
</xs:complexType>
``` |

This type differs from the corresponding test position types used by the other DAL extensions in that it allows one to specify a coordinate reference frame supported by SSA's POS input parameter (see [SSA], section 4.1.1.1).

| ssap:PosParam Metadata Elements | | |
|---|---|---|
| **Element** | **Definition** | |
| long | *Value type:* | floating-point number: `xs:double` |
| | *Semantic Meaning:* | The longitude (e.g. Right Ascension) of the center of the search position in decimal degrees. |
| | *Occurrences:* | required |
| lat | *Value type:* | floating-point number: `xs:double` |
| | *Semantic Meaning:* | The latitude (e.g. Declination) of the center of the search position in decimal degrees. |
| | *Occurrences:* | required |
| refframe | *Value type:* | string: `xs:token` |
| | *Semantic Meaning:* | the coordinate system reference frame name indicating the frame to assume for the given position. If not provided, ICRS is assumed. |
| | *Occurrences:* | optional |

### 3.3.3. ProtoSpectralAccess

The `ssap:ProtoSpectralAccess` type is provided for a special class of SSA services that were historically deployed prior to the publication of the SSA Recommendation [SSA]. An SSA service should describe its support for the protocol using this capability type if it was implemented against an earlier draft of the protocol specification and, therefore, is not expected to be compliant with the actual SSA Recommendation.

This type is defined exactly as the `ssap:SimpleSpectralAccess` type:

| ssap:ProtoSpectralAccess Type Schema Definition |
|---|
| ```
<xs:complexType name="ProtoSpectralAccess" >
  <xs:complexContent >
    <xs:extension base="ssap:SSACapRestriction" >
      <xs:sequence >
        <xs:element name="dataSource" type="ssap:DataSource"
                    minOccurs="1" maxOccurs="unbounded" />
        <xs:element name="creationType" type="ssap:CreationType"
                    minOccurs="1" maxOccurs="unbounded" />
        <xs:element name="supportedFrame" type="ssap:SupportedFrame"
                    minOccurs="1" maxOccurs="unbounded"/>
        <xs:element name="maxSearchRadius" type="xs:double" minOccurs="0" maxOccurs="1" />
        <xs:element name="maxRecords" type="xs:positiveInteger" minOccurs="0" maxOccurs="1" />
        <xs:element name="defaultMaxRecords" type="xs:positiveInteger"
                    minOccurs="0" maxOccurs="1" />
        <xs:element name="maxAperture" type="xs:double" minOccurs="0" maxOccurs="1" />
        <xs:element name="maxFileSize" type="xs:positiveInteger" minOccurs="0" maxOccurs="1" />
        <xs:element name="testQuery" type="ssap:Query" minOccurs="0" maxOccurs="1" />
      </xs:sequence>
    </xs:extension>
  </xs:complexContent>
</xs:complexType>
``` |

Use of this type is intended to be temporary: providers whose SSA service falls into this category are encouraged to up-date the service for compliance with the final SSA Recommendation. A VOResource resource description must not include both a `ssap:SimpleSpectralAccess` capability and a `ssap:ProtoSpectralAccess` capability that describe the same service base URL, as given by the `<interface>`'s `<accessURL>`.

## 3.4. Simple Line Access

This section describes the SLA VOResource metadata extension schema which is used to describe services that comply with the Simple Line Access protocol [SLA].

### 3.4.1. The Schema Namespace

The namespace associated with the SLA extension schema is "http://www.ivoa.net/xml/SLAP/v1.0". The namespace prefix, `slap`, should be used in applications where common use of prefixes improves interoperability (e.g. in the IVOA registries [RI]). Furthermore, we use the `slap` prefix in this document to refer to types defined as part of the SLA extension schema.

### 3.4.2. SimpleLineAccess

The `slap:SimpleLineAccess` type is a `vr:Capability` sub-type that should be used to describe a service's support for the Simple Line Access protocol [SLA]; it is defined as follows:

---

**slap:SLAPCapRestriction Type Schema Definition**

```
<xs:complexType name="SLAPCapRestriction" abstract="true" >
  <xs:complexContent >
    <xs:restriction base="vr:Capability" >
      <xs:sequence >
        <xs:element name="validationLevel" type="vr:Validation" minOccurs="0"
                    maxOccurs="unbounded" />
        <xs:element name="description" type="xs:token" minOccurs="0" />
        <xs:element name="interface" type="vr:Interface" minOccurs="0"
                    maxOccurs="unbounded" />
      </xs:sequence>
      <xs:attribute name="standardID" type="vr:IdentifierURI" use="required"
                    fixed="ivo://ivoa.net/std/SLAP" />
    </xs:restriction>
  </xs:complexContent>
</xs:complexType>

<xs:complexType name="SimpleLineAccess" >
  <xs:complexContent >
    <xs:extension base="slap:SLAPCapRestriction" >
      <xs:sequence >
        <xs:element name="complianceLevel" type="slap:ComplianceLevel" />
        <xs:element name="dataSource" type="slap:DataSource" />
        <xs:element name="maxRecords" type="xs:positiveInteger" minOccurs="0" maxOccurs="1" />
        <xs:element name="testQuery" type="slap:Query" minOccurs="0" maxOccurs="1" />
      </xs:sequence>
    </xs:extension>
  </xs:complexContent>
</xs:complexType>
```

The custom metadata that the `slap:SimpleLineAccess` type provides is given in the table below.

| slap:SimpleLineAccess Extension Metadata Elements | |
|---|---|
| **Element** | **Definition** |
| complianceLevel | *Value type:* string with controlled vocabulary |
| | *Semantic Meaning:* The category indicating the level to which this service instance complies with the SLAP standard. |
| | *Occurrences:* required |
| | *Allowed Values:*    `minimal` The service supports all of the capabilities and features of the SLAP protocol identified as "must" in the specification. |
| |                 `full`   The service supports, at a minimum, all of the capabilities and features of the SLAP protocol identified as "must" or "should" in the specification. |
| | *Comments:* Allowed values are "minimal" and "full". See definitions of allowed values for details. |
| dataSource | *Value type:* string with controlled vocabulary |
| | *Semantic Meaning:* The category specifying where the data accessed by the service originally came from. |
| | *Occurrences:* required |
| | *Allowed Values:*    `observational/astrophysical` Lines observed and identified in real spectra of astrophysical observations by different instrument/projects |
| |        `observational/laboratory`   Lines observed and identified in real spectra of laboratory measurements |
| |                `theoretical`   Servers containing theoretical spectral lines |
| | *Comments:* Allowed values are "observational/astrophysical", "observational/laboratory", "theoretical" |
| maxRecords | *Value type:* positive integer: `xs:positiveInteger` |
| | *Semantic Meaning:* The hard limit on the largest number of records that the query operation will return in a single response. Not providing this value means that there is no effective limit. |
| | *Occurrences:* optional |
| | *Comments:* This does not refer to the total number of records in the catalog but rather maximum number of records the service is capable of returning. A limit that is greater than the number of records available in the archive is equivalent to their being no effective limit. (See also [RM], Sect. 5.2.) |
| testQuery | *Value type:* composite: `slap:Query` |
| | *Semantic Meaning:* A set of queryData parameters that is expected to produce at least one matched record which can be used to test the service. |
| | *Occurrences:* optional |
| | *Comments:* The value should include all parameters required for the test query but should exclude the baseURL and the REQUEST parameter. |

The controlled vocabulary given for the above elements are set by their respective simple types:

---

**slap:ComplianceLevel Type Schema Definition**

```
<xs:simpleType name="ComplianceLevel" >
  <xs:restriction base="xs:token" >
    <xs:enumeration value="minimal" />
    <xs:enumeration value="full" />
  </restriction>
</simpleType>
```

---

**slap:DataSource Type Schema Definition**

```
<xs:simpleType name="DataSource" >
  <xs:restriction base="xs:token" >
    <xs:enumeration value="observational/astrophysical" />
    <xs:enumeration value="observational/laboratory" />
    <xs:enumeration value="theoretical" />
  </restriction>
</simpleType>
```

---

### 3.3.2.1 testQuery and the Query Type

As with the other DAL `vr:capability` types, the `<testQuery>` element is intended to help other VO components (e.g. registries, validation services, services that monitor the VO's operational health--but typically not end users) test that the service is up and operating correctly. It provides a set of legal input parameters that should return a legal response that includes at least matched record. Since this query is intended for testing purposes, the size of the result set should be small.

The `slap:Query` type captures the different components of the query into separate elements, as defined below:

---

**slap:Query Type Schema Definition**

```
<xs:complexType name="Query" >
  <xs:sequence >
    <xs:element name="wavelength" type="slap:WavelengthRange" minOccurs="0" />
    <xs:element name="queryDataCmd" type="xs:string" minOccurs="0" />
  </sequence>
</complexType>
```

The individual sub-elements are defined as follows:

| slap:Query Metadata Elements | |
|---|---|
| **Element** | **Definition** |
| wavelength | *Value type:* composite: slap:WavelengthRange<br><br>*Semantic Meaning:* Spectral range in meters to be used to constrain the query of spectral lines.<br><br>*Occurrences:* optional |
| queryDataCmd | *Value type:* string: `xs:string`<br><br>*Semantic Meaning:* Fully specified queryData test query formatted as an URL argument list in the syntax specified by the SLAP standard. The list must exclude the REQUEST argument which is assumed to be set to "queryData". VERSION may be included if the test query applies to a specific version of the service protocol.<br><br>*Occurrences:* optional<br><br>*Comments:* If queryDataCmd is used to form a query, the default value of WAVELENGTH specified above is not used; if the test query requires WAVELENGTH it should be included directly in queryDataCmd.<br><br>This value must be a string in the form of name=value pairs delimited with ampersands (&). A query may then be formed by appending to the baseURL the request argument, "REQUEST=queryData&", followed by the contents of this element. |

## 3.4.2.2. WavelengthRange

The `slap:WavelengthRange` type is used to encode the `<testQuery>`'s `<wavelength>` element, the range of wavelengths to search.

| slap:WavelengthRange Type Schema Definition |
|---|
| ```
<xs:complexType name="WavelengthRange" >
  <xs:sequence >
    <xs:element name="minWavelength" type="xs:double" minOccurs="0" />
    <xs:element name="maxWavelength" type="xs:double" minOccurs="0" />
  </xs:sequence>
</xs:complexType>
``` |

| slap:WavelengthRange Metadata Elements | |
|---|---|
| **Element** | **Definition** |
| minWavelength | *Value type:* floating-point number: `xs:double`<br><br>*Semantic Meaning:* Minimum wavelength in meters to be used to constrain the query of spectral lines<br><br>*Occurrences:* optional |
| maxWavelength | *Value type:* floating-point number: `xs:double`<br><br>*Semantic Meaning:* Maximum wavelength in meters to be used to constrain the query of spectral lines<br><br>*Occurrences:* optional |

# Appendix A: The XML Schemas

## A.1. The ConeSearch XML Schema

### The Complete ConeSearch Schema

```
<?xml version="1.0" encoding="UTF-8"?>
<xs:schema xmlns:xs="http://www.w3.org/2001/XMLSchema"
           xmlns:vr="http://www.ivoa.net/xml/VOResource/v1.0"
           xmlns:cs="http://www.ivoa.net/xml/ConeSearch/v1.0"
           xmlns:vm="http://www.ivoa.net/xml/VOMetadata/v0.1"
           targetNamespace="http://www.ivoa.net/xml/ConeSearch/v1.0"
           elementFormDefault="unqualified" attributeFormDefault="unqualified"
           version="1.0">

    <xs:annotation>
        <xs:appinfo>
          <vm:schemaName>ConeSearch</vm:schemaName>
          <vm:schemaPrefix>xs</vm:schemaPrefix>
          <vm:targetPrefix>cs</vm:targetPrefix>
        </xs:appinfo>
        <xs:documentation>
          Implementation of an XML Schema describing a Cone Search Service.
          Based on "Simple Cone Search: a First Guide for Data Curators",
          http://www.us-vo.org/metadata/conesearch/.
        </xs:documentation>
        <xs:documentation>
          This schema defines a Capability type that is specific to Cone
          Search.  A service that has ConeSearch capability should be
          registered as a TabularSkyService.  The ConeSearch capability
          element must include a ParamHTTP interface.
        </xs:documentation>
    </xs:annotation>

    <xs:import namespace="http://www.ivoa.net/xml/VOResource/v1.0"
               schemaLocation="http://www.ivoa.net/xml/VOResource/v1.0"/>

    <xs:complexType name="CSCapRestriction" abstract="true">
        <xs:annotation>
           <xs:documentation>
              an abstract capability that fixes the standardID to the
              IVOA ID for the ConeSearch standard.
           </xs:documentation>
           <xs:documentation>
              See vr:Capability for documentation on inherited children.
           </xs:documentation>
        </xs:annotation>
        <xs:complexContent>
            <xs:restriction base="vr:Capability">
                <xs:sequence>
                    <xs:element name="validationLevel" type="vr:Validation"
                                minOccurs="0" maxOccurs="unbounded"/>
                    <xs:element name="description" type="xs:token"
                                minOccurs="0"/>
                    <xs:element name="interface" type="vr:Interface"
                                minOccurs="0" maxOccurs="unbounded"/>
                </xs:sequence>
                <xs:attribute name="standardID" type="vr:IdentifierURI"
                              use="required" fixed="ivo://ivoa.net/std/ConeSearch"/>
            </xs:restriction>
        </xs:complexContent>
    </xs:complexType>

    <xs:complexType name="ConeSearch">
        <xs:annotation>
           <xs:documentation>
              The capabilities of a Cone Search implementation.
           </xs:documentation>
        </xs:annotation>

        <xs:complexContent>
```

```xml
            <xs:extension base="cs:CSCapRestriction">
                <xs:sequence>
                    <xs:element name="maxSR" type="xs:float"
                                minOccurs="0" maxOccurs="1">
                        <xs:annotation>
                            <xs:documentation>
                                The largest search radius, in degrees, that will be
                                accepted by the service without returning an error
                                condition.  Not providing this element or
                                specifying a value of 180 indicates that there
                                is no restriction.
                            </xs:documentation>
                            <xs:documentation>
                                Not providing a value is the prefered way to indicate
                                that there is no restriction.
                            </xs:documentation>
                        </xs:annotation>
                    </xs:element>

                    <xs:element name="maxRecords" type="xs:positiveInteger"
                                minOccurs="0" maxOccurs="1">
                        <xs:annotation>
                            <xs:documentation>
                                The largest number of records that the service will
                                return.  Not providing this value means that
                                there is no effective limit.
                            </xs:documentation>
                            <xs:documentation>
                                This does not refer to the total number of records in
                                the catalog but rather maximum number of records the
                                service is capable of returning.  A limit that is
                                greater than the number of records available in the
                                archive is equivalent to their being no effective
                                limit.  (See RM, Hanisch 2007.)
                            </xs:documentation>
                        </xs:annotation>
                    </xs:element>

                    <xs:element name="verbosity" type="xs:boolean">
                        <xs:annotation>
                            <xs:documentation>
                                True if the service supports the VERB keyword;
                                false, otherwise.
                            </xs:documentation>
                        </xs:annotation>
                    </xs:element>

                    <xs:element name="testQuery" type="cs:Query"
                                minOccurs="0" maxOccurs="1">
                        <xs:annotation>
                            <xs:documentation>
                                A query that will result in at least on
                                matched record that can be used to test the
                                service.
                            </xs:documentation>
                        </xs:annotation>
                    </xs:element>

                </xs:sequence>
            </xs:extension>
        </xs:complexContent>
    </xs:complexType>

    <xs:complexType name="Query">
        <xs:annotation>
            <xs:documentation>
                A query to be sent to the service
            </xs:documentation>
        </xs:annotation>

        <xs:sequence>
            <xs:element name="ra" type="xs:double">
                <xs:annotation>
```

```
                    <xs:documentation>
                        the right ascension of the search cone's center in
                        decimal degrees.
                    </xs:documentation>
                </xs:annotation>
            </xs:element>

            <xs:element name="dec" type="xs:double">
                <xs:annotation>
                    <xs:documentation>
                        the declination of the search cone's center in
                        decimal degrees.
                    </xs:documentation>
                </xs:annotation>
            </xs:element>

            <xs:element name="sr" type="xs:double">
                <xs:annotation>
                    <xs:documentation>
                        the radius of the search cone in decimal degrees.
                    </xs:documentation>
                </xs:annotation>
            </xs:element>

            <xs:element name="verb" type="xs:positiveInteger" minOccurs="0">
                <xs:annotation>
                    <xs:documentation>
                        the verbosity level to use where 1 means the bare
                        minimum set of columns and 3 means the full set of
                        available columns.
                    </xs:documentation>
                </xs:annotation>
            </xs:element>

            <xs:element name="catalog" type="xs:string" minOccurs="0">
                <xs:annotation>
                    <xs:documentation>
                        the catalog to query.
                    </xs:documentation>
                    <xs:documentation>
                        When the service can access more than one catalog,
                        this input parameter, if available, is used to
                        indicate which service to access.
                    </xs:documentation>
                </xs:annotation>
            </xs:element>

            <xs:element name="extras" type="xs:string" minOccurs="0">
                <xs:annotation>
                    <xs:documentation>
                        any extra (non-standard) parameters that must be
                        provided (apart from what is part of base URL given
                        by the accessURL element).
                    </xs:documentation>
                    <xs:documentation>
                        this value should be in the form of name=value
                        pairs delimited with ampersands (&).
                    </xs:documentation>
                </xs:annotation>
            </xs:element>

        </xs:sequence>
    </xs:complexType>

</xs:schema>
```

## A.2. The SIA XML Schema

### The Complete SIA Schema

```xml
<?xml version="1.0" encoding="UTF-8"?>
<xs:schema xmlns:xs="http://www.w3.org/2001/XMLSchema"
           xmlns:vr="http://www.ivoa.net/xml/VOResource/v1.0"
           xmlns:sia="http://www.ivoa.net/xml/SIA/v1.1"
           xmlns:vm="http://www.ivoa.net/xml/VOMetadata/v0.1"
           targetNamespace="http://www.ivoa.net/xml/SIA/v1.1"
           elementFormDefault="unqualified" attributeFormDefault="unqualified"
           version="1.1">

    <xs:annotation>
        <xs:appinfo>
          <vm:schemaName>SIA</vm:schemaName>
          <vm:schemaPrefix>xs</vm:schemaPrefix>
          <vm:targetPrefix>sia</vm:targetPrefix>
        </xs:appinfo>
        <xs:documentation>
          Implementation of an XML Schema describing a Simple Image
          Access Service.   Based on "The Simple Image Access Specification"
          by Doug Tody et al.
        </xs:documentation>
    </xs:annotation>

    <xs:import namespace="http://www.ivoa.net/xml/VOResource/v1.0"
               schemaLocation="http://www.ivoa.net/xml/VOResource/v1.0"/>

    <xs:complexType name="SIACapRestriction" abstract="true">
        <xs:annotation>
            <xs:documentation>
                an abstract capability that fixes the standardID to the
                IVOA ID for the ConeSearch standard.
            </xs:documentation>
            <xs:documentation>
                See vr:Capability for documentation on inherited children.
            </xs:documentation>
        </xs:annotation>
        <xs:complexContent>
            <xs:restriction base="vr:Capability">
                <xs:sequence>
                    <xs:element name="validationLevel" type="vr:Validation"
                                minOccurs="0" maxOccurs="unbounded"/>
                    <xs:element name="description" type="xs:token"
                                minOccurs="0"/>
                    <xs:element name="interface" type="vr:Interface"
                                minOccurs="0" maxOccurs="unbounded"/>
                </xs:sequence>
                <xs:attribute name="standardID" type="vr:IdentifierURI"
                              use="required" fixed="ivo://ivoa.net/std/SIA"/>
            </xs:restriction>
        </xs:complexContent>
    </xs:complexType>

    <xs:complexType name="SimpleImageAccess">
        <xs:annotation>
            <xs:documentation>
                The capabilities of an SIA implementation.
            </xs:documentation>
            <xs:documentation>
                Editor's Notes:
                    *   This is a prototype definition to
                        illustrate how to extend the schema to a specific
                        standard interface.
                    *   Staging information is not included yet.
            </xs:documentation>
        </xs:annotation>

        <xs:complexContent>
            <xs:extension base="sia:SIACapRestriction">
                <xs:sequence>
```

```xml
        <xs:element name="imageServiceType" type="sia:ImageServiceType">
            <xs:annotation>
                <xs:documentation>
                 The class of image service: Cutout, Mosaic, Atlas, Pointed
                </xs:documentation>
            </xs:annotation>
        </xs:element>

        <xs:element name="maxQueryRegionSize" type="sia:SkySize"
                    minOccurs="0" maxOccurs="1">
            <xs:annotation>
                <xs:documentation>
                    The maximum image query region size, expressed in
                    decimal degrees.  Not providing this element or
                    specifying a value of 360 degrees indicates that
                    there is no limit and the entire data collection
                    (entire sky) can be queried.
                </xs:documentation>
                <xs:documentation>
                    Not providing a value is the prefered way to indicate
                    that there is no limit.
                </xs:documentation>
            </xs:annotation>
        </xs:element>

        <xs:element name="maxImageExtent" type="sia:SkySize"
                    minOccurs="0" maxOccurs="1">
            <xs:annotation>
                <xs:documentation>
                    An upper bound on a region of the sky that can
                    be covered by returned images.  That is, no image
                    returned by this service will cover more than
                    this limit.  Not providing this element or
                    specifying a value of 360 degrees indicates that
                    there is no fundamental limit to the region covered
                    by a returned image.
                </xs:documentation>
                <xs:documentation>
                    When the imageServiceType is "Cutout" or "Mosaic",
                    this represents the largest area that can be requested.
                    In this case, the "no limit" value means that all-sky
                    images can be requested.  When the type is "Atlas" or
                    "Pointed", it should be a region that most closely
                    encloses largest images in the archive, and the "no
                    limit" value means that the archive contains all-sky
                    (or nearly so) images.
                </xs:documentation>
                <xs:documentation>
                    Not providing a value is the prefered way to indicate
                    that there is no limit.
                </xs:documentation>
            </xs:annotation>
        </xs:element>

        <xs:element name="maxImageSize" type="xs:positiveInteger"
                    minOccurs="0" maxOccurs="1">
            <xs:annotation>
                <xs:documentation>
                    A measure of the largest image the service
                    can produce given as the maximum number of
                    pixels along the first or second axes.
                    Not providing a value indicates that there is
                    no effective limit to the size of the images
                    that can be returned.
                </xs:documentation>
                <xs:documentation>
                    This is primarily relevant when the imageServiceType
                    is "Cutout" or "Mosaic", indicating the largest
                    image that can be created.  When the imageServiceType
                    is "Atlas" or "Pointed", this should be specified only
                    when there are static images in the archive that can
                    be searched for but not returned because they are
                    too big.
```

```
                    </xs:documentation>
                    <xs:documentation>
                        When a service is more fundamentally limited
                        by the total number of pixels in the image, this
                        value should be set to the square-root of that
                        number.  This number will then represent a
                        lower limit on the maximum length of a side.
                    </xs:documentation>
                </xs:annotation>
            </xs:element>

            <xs:element name="maxFileSize" type="xs:positiveInteger"
                        minOccurs="0" maxOccurs="1">
                <xs:annotation>
                    <xs:documentation>
                        The maximum image file size in bytes.  Not providing
                        a value indicates that there is no effective limit
                        the size of files that can be returned.
                    </xs:documentation>
                    <xs:documentation>
                        This is primarily relevant when the imageServiceType
                        is "Cutout" or "Mosaic", indicating the largest
                        files that can be created.  When the imageServiceType
                        is "Atlas" or "Pointed", this should be specified only
                        when there are static images in the archive that can
                        be searched for but not returned because they are
                        too big.
                    </xs:documentation>
                </xs:annotation>
            </xs:element>

            <xs:element name="maxRecords" type="xs:positiveInteger"
                        minOccurs="0" maxOccurs="1">
                <xs:annotation>
                    <xs:documentation>
                        The largest number of records that the Image Query web
                        method will return. Not providing this value means that
                        there is no effective limit.
                    </xs:documentation>
                    <xs:documentation>
                        This does not refer to the total number of images in
                        the archive but rather maximum number of records the
                        service is capable of returning.  A limit that is
                        greater than the number of images available in the
                        archive is equivalent to their being no effective
                        limit.  (See RM, Hanisch 2007.)
                    </xs:documentation>
                </xs:annotation>
            </xs:element>

            <xs:element name="testQuery" type="sia:Query"
                        minOccurs="0" maxOccurs="1">
                <xs:annotation>
                    <xs:documentation>
                        a set of query parameters that is expected
                        to produce at least one matched record which
                        can be used to test the service.
                    </xs:documentation>
                </xs:annotation>
            </xs:element>

        </xs:sequence>
      </xs:extension>
    </xs:complexContent>
</xs:complexType>

<xs:complexType name="SkySize">
    <xs:sequence>
        <xs:element name="long" type="xs:double">
            <xs:annotation>
                <xs:documentation>
                    The maximum size in the longitude (R.A.) direction
                    given in degrees
```

```xml
                </xs:documentation>
              </xs:annotation>
            </xs:element>
            <xs:element name="lat" type="xs:double">
              <xs:annotation>
                <xs:documentation>
                    The maximum size in the latitude (Dec.) direction
                    given in degrees
                </xs:documentation>
              </xs:annotation>
            </xs:element>
        </xs:sequence>
    </xs:complexType>

    <xs:complexType name="SkyPos">
        <xs:sequence>
            <xs:element name="long" type="xs:double">
              <xs:annotation>
                <xs:documentation>
                    The sky position in the longitude (R.A.) direction
                </xs:documentation>
              </xs:annotation>
            </xs:element>
            <xs:element name="lat" type="xs:double">
              <xs:annotation>
                <xs:documentation>
                    The sky position in the latitude (Dec.) direction
                </xs:documentation>
              </xs:annotation>
            </xs:element>
        </xs:sequence>
    </xs:complexType>

    <xs:simpleType name="ImageServiceType">
        <xs:annotation>
          <xs:documentation>
         The class of image service: Cutout, Mosaic, Atlas, Pointed
          </xs:documentation>
        </xs:annotation>

        <xs:restriction base="xs:token">

            <xs:enumeration value="Cutout">
              <xs:annotation>
                <xs:documentation>
         This is a service which extracts or "cuts out" rectangular
         regions of some larger image, returning an image of the
         requested size to the client. Such images are usually drawn
         from a database or a collection of survey images that cover
         some large portion of the sky. To be considered a cutout
         service, the returned image should closely approximate (or at
         least not exceed) the size of the requested region; however,
         a cutout service will not normally resample (rescale or
         reproject) the pixel data. A cutout service may mosaic image
         segments to cover a large region but is still considered a
         cutout service if it does not resample the data.  Image
         cutout services are fast and avoid image degredation due to
         resampling.
                </xs:documentation>
              </xs:annotation>
            </xs:enumeration>

            <xs:enumeration value="Mosaic">
              <xs:annotation>
                <xs:documentation>
                    This service is similar to the image cutout service
                    but adds the capability to compute an image of the
                    size, scale, and projection specified by the
                    client. Mosaic services include services which
                    resample and reproject existing image data, as well
                    as services which generate pixels from some more
                    fundamental dataset, e.g., a high energy event list
                    or a radio astronomy measurement set. Image mosaics
```

```
                        can be expensive to generate for large regions but
                        they make it easier for the client to overlay image
                        data from different sources. Image mosaicing
                        services which resample already pixelated data will
                        degrade the data slightly, unlike the simpler cutout
                        service which returns the data unchanged.
                    </xs:documentation>
                </xs:annotation>
            </xs:enumeration>

            <xs:enumeration value="Atlas">
                <xs:annotation>
                    <xs:documentation>
                        This category of service provides access to
                        pre-computed images that make up a survey of some
                        large portion of the sky. The service, however, is
                        not capable of dynamically cutting out requested
                        regions, and the size of atlas images is
                        predetermined by the survey. Atlas images may range
                        in size from small cutouts of extended objects to
                        large calibrated survey data frames.
                    </xs:documentation>
                </xs:annotation>
            </xs:enumeration>

            <xs:enumeration value="Pointed">
                <xs:annotation>
                    <xs:documentation>
                        This category of service provides access to
                        collections of images of many small, "pointed"
                        regions of the sky. "Pointed" images normally focus
                        on specific sources in the sky as opposed to being
                        part of a sky survey. This type of service usually
                        applies to instrumental archives from observatories
                        with guest observer programs (e.g., the HST archive)
                        and other general purpose image archives (e.g., the
                        ADIL). If a service provides access to both survey
                        and pointed images, then it should be considered a
                        Pointed Image Archive for the purposes of this
                        specification; if a differentiation between the
                        types of data is desired the pointed and survey data
                        collections should be registered as separate image
                        services.
                    </xs:documentation>
                </xs:annotation>
            </xs:enumeration>

    </xs:restriction>
</xs:simpleType>

<xs:complexType name="Query">
    <xs:annotation>
        <xs:documentation>
            A query to be sent to the service
        </xs:documentation>
    </xs:annotation>

    <xs:sequence>
        <xs:element name="pos" type="sia:SkyPos" minOccurs="0">
            <xs:annotation>
                <xs:documentation>
                    the center position of the rectangular region that
                    should be used as part of the query to the SIA service.
                </xs:documentation>
            </xs:annotation>
        </xs:element>

        <xs:element name="size" type="sia:SkySize" minOccurs="0">
            <xs:annotation>
                <xs:documentation>
                    the rectangular size of the region that should be
                    used as part of the query to the SIA service.
                </xs:documentation>
```

```
                </xs:annotation>
            </xs:element>

            <xs:element name="verb" type="xs:positiveInteger" minOccurs="0">
                <xs:annotation>
                    <xs:documentation>
                        the verbosity level to use where 0 means the bare
                        minimum set of columns and 3 means the full set of
                        available columns.
                    </xs:documentation>
                </xs:annotation>
            </xs:element>

            <xs:element name="extras" type="xs:string" minOccurs="0">
                <xs:annotation>
                    <xs:documentation>
                        any extra (particularly non-standard) parameters that must
                        be provided (apart from what is part of base URL given by
                        the accessURL element).
                    </xs:documentation>
                    <xs:documentation>
                        this value should be in the form of name=value
                        pairs delimited with apersands (&).
                    </xs:documentation>
                </xs:annotation>
            </xs:element>

        </xs:sequence>

    </xs:complexType>

</xs:schema>
```

## A.3 The SSA XML Schema

**The Complete SSA Schema**

```
<?xml version="1.0" encoding="UTF-8"?>
<xs:schema xmlns:xs="http://www.w3.org/2001/XMLSchema"
           xmlns:vr="http://www.ivoa.net/xml/VOResource/v1.0"
           xmlns:ssap="http://www.ivoa.net/xml/SSA/v1.1"
           xmlns:vm="http://www.ivoa.net/xml/VOMetadata/v0.1"
           targetNamespace="http://www.ivoa.net/xml/SSA/v1.1"
           elementFormDefault="unqualified" attributeFormDefault="unqualified"
           version="1.1">

    <xs:annotation>
        <xs:appinfo>
            <vm:schemaName>SSA</vm:schemaName>
            <vm:schemaPrefix>xs</vm:schemaPrefix>
            <vm:targetPrefix>ssap</vm:targetPrefix>
        </xs:appinfo>
        <xs:documentation>
            XML Schema used to describe the capabilities of a service instance
            conforming to the Simple Spectral Access (SSA) protocol.
        </xs:documentation>
    </xs:annotation>

    <xs:import namespace="http://www.ivoa.net/xml/VOResource/v1.0"
               schemaLocation="http://www.ivoa.net/xml/VOResource/v1.0"/>

    <!-- Set the Capability standardID to indicate the SSA protocol. -->
    <xs:complexType name="SSACapRestriction" abstract="true">
        <xs:annotation>
            <xs:documentation>
                An abstract capability that fixes the standardID to the
                IVOA ID for the SSA standard.
            </xs:documentation>
            <xs:documentation>
                See vr:Capability for documentation on inherited children.
            </xs:documentation>
```

```
            </xs:annotation>

        <xs:complexContent>
            <xs:restriction base="vr:Capability">
                <xs:sequence>
                    <xs:element name="validationLevel" type="vr:Validation"
                                minOccurs="0" maxOccurs="unbounded"/>
                    <xs:element name="description" type="xs:token"
                                minOccurs="0"/>
                    <xs:element name="interface" type="vr:Interface"
                                minOccurs="0" maxOccurs="unbounded"/>
                </xs:sequence>
                <xs:attribute name="standardID" type="vr:IdentifierURI"
                              use="required" fixed="ivo://ivoa.net/std/SSA"/>
            </xs:restriction>
        </xs:complexContent>
</xs:complexType>

<!-- SSA Capabilities -->
<xs:complexType name="SimpleSpectralAccess">
    <xs:annotation>
        <xs:documentation>
            The capabilities of an SSA service implementation.
        </xs:documentation>
    </xs:annotation>

    <xs:complexContent>
        <xs:extension base="ssap:SSACapRestriction">
            <xs:sequence>

                <xs:element name="complianceLevel" type="ssap:ComplianceLevel">
                    <xs:annotation>
                        <xs:documentation>
                            The category indicating the level to which
                            this instance complies with the SSA standard.
                        </xs:documentation>
                        <xs:documentation>
                            Allowed values are "query", "minimal", and "full".
                            See definitions of allowed values for details.
                        </xs:documentation>
                    </xs:annotation>
                </xs:element>

                <xs:element name="dataSource" type="ssap:DataSource"
                            minOccurs="1" maxOccurs="unbounded">
                    <xs:annotation>
                        <xs:documentation>
                            The category specifying where the data originally
                            came from.
                        </xs:documentation>
                        <xs:documentation>
                            Allowed values are "survey", "pointed", "custom",
                            "theory", "artificial"
                        </xs:documentation>
                    </xs:annotation>
                </xs:element>

                <xs:element name="creationType" type="ssap:CreationType"
                            minOccurs="1" maxOccurs="unbounded">
                    <xs:annotation>
                        <xs:documentation>
                            The category that describes the process used to
                            produce the dataset.
                        </xs:documentation>
                        <xs:documentation>
                            Typically this describes only the processing
                            performed by the data service, but it could
                            describe some additional earlier processing as
                            well, e.g., if data is partially precomputed.
                        </xs:documentation>
                        <xs:documentation>
                            Allowed values are "archival", "cutout", "filtered",
                            "mosaic", "projection", "spectralExtraction",
```

```xml
                        "catalogExtraction"
                    </xs:documentation>
                </xs:annotation>
            </xs:element>

            <xs:element name="supportedFrame" type="ssap:SupportedFrame"
                        minOccurs="1" maxOccurs="unbounded">
                <xs:annotation>
                    <xs:documentation>
                        The STC name for a world coordinate system
                        frame supported by this service.
                    </xs:documentation>
                    <xs:documentation>
                        At least one recognized value must be listed.
                        With SSA v1.1, ICRS must be supported; thus,
                        this list must include at least this value.
                    </xs:documentation>
                </xs:annotation>
            </xs:element>

            <xs:element name="maxSearchRadius" type="xs:double"
                        minOccurs="0" maxOccurs="1">
                <xs:annotation>
                    <xs:documentation>
                        The largest search radius, in degrees, that will be
                        accepted by the service without returning an error
                        condition.  Not providing this element or
                        specifying a value of 180 indicates that there
                        is no restriction.
                    </xs:documentation>
                    <xs:documentation>
                        Not providing a value is the prefered way to indicate
                        that there is no restriction.
                    </xs:documentation>
                </xs:annotation>
            </xs:element>

            <xs:element name="maxRecords" type="xs:positiveInteger"
                        minOccurs="0" maxOccurs="1">
                <xs:annotation>
                    <xs:documentation>
                        The hard limit on the largest number of records that
                        the query operation will return in a single response.
                        Not providing this value means that there is no
                        effective limit.
                    </xs:documentation>
                    <xs:documentation>
                        This does not refer to the total number of spectra in
                        the archive but rather maximum number of records the
                        service is capable of returning.  A limit that is
                        greater than the number of spectra available in the
                        archive is equivalent to their being no effective
                        limit.  (See RM, Hanisch 2007.)
                    </xs:documentation>
                </xs:annotation>
            </xs:element>

            <xs:element name="defaultMaxRecords" type="xs:positiveInteger"
                        minOccurs="0" maxOccurs="1">
                <xs:annotation>
                    <xs:documentation>
                        The largest number of records that the service will
                        return when the MAXREC parameter not specified
                        in the query input.  Not providing a value means
                        that the hard limit implied by maxRecords will be
                        the default limit.
                    </xs:documentation>
                </xs:annotation>
            </xs:element>

            <xs:element name="maxAperture" type="xs:double"
                        minOccurs="0" maxOccurs="1">
                <xs:annotation>
```

```
                    <xs:documentation>
                        The largest aperture that can be supported upon
                        request via the APERTURE input parameter by a
                        service that supports the spectral extraction
                        creation method.  A value of 180 or not providing
                        a value means there is no theoretical limit.
                    </xs:documentation>
                    <xs:documentation>
                        Not providing a value is the preferred way to
                        indicate that there is no limit.
                    </xs:documentation>
                </xs:annotation>
            </xs:element>

            <xs:element name="maxFileSize" type="xs:positiveInteger"
                        minOccurs="0" maxOccurs="1">
                <xs:annotation>
                    <xs:documentation>
                        The maximum spectrum file size in bytes that will
                        be returned.  Not providing
                        a value indicates that there is no effective limit
                        the size of files that can be returned.
                    </xs:documentation>
                    <xs:documentation>
                        This is primarily relevant when spectra are created
                        on the fly (see creationType).  If the service
                        provides access to static spectra, this should only
                        be specified if there are spectra in the archive that
                        can be searched for but not returned because they are
                        too big.
                    </xs:documentation>
                </xs:annotation>
            </xs:element>

            <xs:element name="testQuery" type="ssap:Query"
                        minOccurs="0" maxOccurs="1">
                <xs:annotation>
                    <xs:documentation>
                        a set of query parameters that is expected to
                        produce at least one matched record which can be
                        used to test the service.
                    </xs:documentation>
                </xs:annotation>
            </xs:element>

        </xs:sequence>
    </xs:extension>
  </xs:complexContent>
</xs:complexType>

<xs:simpleType name="ComplianceLevel">
    <xs:annotation>
        <xs:documentation>
            The allowed values for indicating the level at which a
            service instance complies with the SSA standard.
        </xs:documentation>
    </xs:annotation>

    <xs:restriction base="xs:token">
        <xs:enumeration value="query">
            <xs:annotation>
                <xs:documentation>
                    The service supports all of the capabilities and features
                    of the SSA protocol identified as "must" in the
                    specification, except that it does not support returning
                    data in at least one SSA-compliant format.
                </xs:documentation>
                <xs:documentation>
                    This level represents the lowest level of compliance.
                </xs:documentation>
            </xs:annotation>
        </xs:enumeration>
```

```xml
            <xs:enumeration value="minimal">
                <xs:annotation>
                    <xs:documentation>
                        The service supports all of the capabilities and features
                        of the SSA protocol identified as "must" in the
                        specification.
                    </xs:documentation>
                    <xs:documentation>
                        In brief, this includes:
                        *  implementing the GET interface,
                        *  support the parameters POS, SIZE, TOME, BAND, and
                            FORMAT
                        *  includes all mandatory metadata fields in query
                            response
                        *  supports getData method retrieval in at least one
                            SSA-compliant format
                        *  supports the "FORMAT=METADATA" metadata query.
                    </xs:documentation>
                    <xs:documentation>
                        This level represents the middle level of compliance.
                    </xs:documentation>
                </xs:annotation>
            </xs:enumeration>

            <xs:enumeration value="full">
                <xs:annotation>
                    <xs:documentation>
                        The service supports all of the capabilities and features
                        of the SSA protocol identified as "must" or "should" in the
                        specification.
                    </xs:documentation>
                    <xs:documentation>
                        This level represents the highest level of compliance.
                    </xs:documentation>
                </xs:annotation>
            </xs:enumeration>
        </xs:restriction>
    </xs:simpleType>

    <xs:simpleType name="DataSource">
        <xs:annotation>
            <xs:documentation>
                The defined categories that specify where the spectral data
                originally came from.
            </xs:documentation>
        </xs:annotation>

        <xs:restriction base="xs:token">
            <xs:enumeration value="survey">
                <xs:annotation>
                    <xs:documentation>
                        A survey dataset, which typically covers some region of
                        observational parameter space in a uniform fashion, with
                        as complete as possible coverage in the region of parameter
                        space observed.
                    </xs:documentation>
                </xs:annotation>
            </xs:enumeration>

            <xs:enumeration value="pointed">
                <xs:annotation>
                    <xs:documentation>
                        A pointed observation of a particular astronomical object
                        or field.
                    </xs:documentation>
                    <xs:documentation>
                        Typically, these are instrumental observations taken as
                        part of some PI observing program. The data quality and
                        characteristics may be variable, but the observations of
                        a particular object or field may be more extensive than
                        for a survey.
                    </xs:documentation>
                </xs:annotation>
            </xs:enumeration>
```

```xml
                        </xs:enumeration>

                        <xs:enumeration value="custom">
                            <xs:annotation>
                                <xs:documentation>
                                    Data which has been custom processed, e.g., as part of
                                    a specific research project.
                                </xs:documentation>
                            </xs:annotation>
                        </xs:enumeration>

                        <xs:enumeration value="theory">
                            <xs:annotation>
                                <xs:documentation>
                                    Theory data, or any data generated from a theoretical
                                    model, for example a synthetic spectrum.
                                </xs:documentation>
                            </xs:annotation>
                        </xs:enumeration>

                        <xs:enumeration value="artificial">
                            <xs:annotation>
                                <xs:documentation>
                                    Artificial or simulated data.
                                </xs:documentation>
                                <xs:documentation>
                                    This is similar to theory data but need not be based on
                                    a physical model, and is often used for testing purposes.
                                </xs:documentation>
                            </xs:annotation>
                        </xs:enumeration>

                    </xs:restriction>
                </xs:simpleType>

                <xs:simpleType name="CreationType">
                    <xs:restriction base="xs:token">
                        <xs:enumeration value="archival">
                            <xs:annotation>
                                <xs:documentation>
                                    The entire archival or project dataset is returned.
                                    Transformations such as metadata or data model mediation
                                    or format conversions may take place, but the content of
                                    the dataset is not substantially modified (e.g., all the
                                    data is returned and the sample values are not modified).
                                </xs:documentation>
                            </xs:annotation>
                        </xs:enumeration>

                        <xs:enumeration value="cutout">
                            <xs:annotation>
                                <xs:documentation>
                                    The dataset is subsetted in some region of parameter
                                    space to produce a subset dataset. Sample values are not
                                    modified, e.g., cutouts could be recombined to reconstitute
                                    the original dataset.
                                </xs:documentation>
                            </xs:annotation>
                        </xs:enumeration>

                        <xs:enumeration value="filtered">
                            <xs:annotation>
                                <xs:documentation>
                                    The data is filtered in some fashion to exclude portions
                                    of the dataset, e.g., passing only data in selected regions
                                    along a measurement axis, or processing the data in a way
                                    which recomputes the sample values, e.g., due to
                                    interpolation or flux transformation.
                                </xs:documentation>
                                <xs:documentation>
                                    Filtering is often
                                    combined with other forms of processing, e.g., projection.
                                </xs:documentation>
```

```
              </xs:annotation>
            </xs:enumeration>

            <xs:enumeration value="mosaic">
              <xs:annotation>
                <xs:documentation>
                  Data from multiple non- or partially-overlapping datasets
                  are combined to produce a new dataset.
                </xs:documentation>
              </xs:annotation>
            </xs:enumeration>

            <xs:enumeration value="projection">
              <xs:annotation>
                <xs:documentation>
                  Data is geometrically warped or dimensionally reduced by
                  projecting through a multidimensional dataset.
                </xs:documentation>
              </xs:annotation>
            </xs:enumeration>

            <xs:enumeration value="spectralExtraction">
              <xs:annotation>
                <xs:documentation>
                  Extraction of a spectrum from another dataset, e.g.,
                  extraction of a spectrum from a spectral data cube
                  through a simulated aperture.
                </xs:documentation>
              </xs:annotation>
            </xs:enumeration>

            <xs:enumeration value="catalogExtraction">
              <xs:annotation>
                <xs:documentation>
                  Extraction of a catalog of some form from another dataset,
                  e.g., extraction of a source catalog from an image, or
                  extraction of a line list catalog from a spectrum (not
                  valid for a SSA service).
                </xs:documentation>
              </xs:annotation>
            </xs:enumeration>

    </xs:restriction>
</xs:simpleType>

<xs:simpleType name="SupportedFrame">
    <xs:annotation>
      <xs:documentation>
        A controlled list of space-time reference frame names
        defined from Table 3 (S 4.4.1.2.3) of the STC spec.
      </xs:documentation>
    </xs:annotation>

    <xs:restriction base="xs:token">
        <xs:enumeration value="FK4">
          <xs:annotation>
            <xs:documentation>
              the Fundemental Katalog, system 4, frame; Besselian
            </xs:documentation>
            <xs:documentation>
              Spectrum data files will require a specification of
              the equinox; if it is not expressed explicitly,
              B1950.0 should be assumed.
            </xs:documentation>
          </xs:annotation>
        </xs:enumeration>

        <xs:enumeration value="FK5">
          <xs:annotation>
            <xs:documentation>
              the Fundemental Katalog, system 5, frame; Julien
            </xs:documentation>
            <xs:documentation>
```

```
                    Spectrum data files will require a specification of
                    the equinox; if it is not expressed explicitly,
                    J2000.0 should be assumed.
                </xs:documentation>
            </xs:annotation>
        </xs:enumeration>

        <xs:enumeration value="ECLIPTIC">
            <xs:annotation>
                <xs:documentation>
                    Ecliptic coordinates
                </xs:documentation>
            </xs:annotation>
        </xs:enumeration>

        <xs:enumeration value="ICRS">
            <xs:annotation>
                <xs:documentation>
                    International Celestial Reference System
                </xs:documentation>
            </xs:annotation>
        </xs:enumeration>

        <xs:enumeration value="GALACTIC_I">
            <xs:annotation>
                <xs:documentation>
                    old Galactic coordinates
                </xs:documentation>
            </xs:annotation>
        </xs:enumeration>

        <xs:enumeration value="GALACTIC_II">
            <xs:annotation>
                <xs:documentation>
                    old Galactic coordinates
                </xs:documentation>
            </xs:annotation>
        </xs:enumeration>

        <xs:enumeration value="SUPER_GALACTIC">
            <xs:annotation>
                <xs:documentation>
                    Super-galactic coordinates with the north pole at
                    GALACTIC_II (47.37, +6.32) and the origin at
                    GALACTIC_II (137.37, 0).
                </xs:documentation>
            </xs:annotation>
        </xs:enumeration>

        <xs:enumeration value="AZ_EL">
            <xs:annotation>
                <xs:documentation>
                    The local azimuth and elevation frame where azimuth
                    increases from north through east.
                </xs:documentation>
            </xs:annotation>
        </xs:enumeration>

        <xs:enumeration value="BODY">
            <xs:annotation>
                <xs:documentation>
                    A generic solar system body-centered coordinate frame
                </xs:documentation>
                <xs:documentation>
                    If applicable, queries against this system should assume a
                    default magnitude value or range in the absence of an
                    applicable (non-standard) query constraint.  Service
                    providers are encouraged to document the such assumptions
                    in the resource or capability description.
                </xs:documentation>
            </xs:annotation>
        </xs:enumeration>
```

```xml
<xs:enumeration value="GEO_C">
    <xs:annotation>
        <xs:documentation>
            3D Geographic (geocentric) coordinates where the magnitude
            is expressed as a geocentric distance
        </xs:documentation>
        <xs:documentation>
            If applicable, queries against this system should assume a
            default magnitude value or range in the absence of an
            applicable (non-standard) query constraint.  Service
            providers are encouraged to document the such assumptions
            in the resource or capability description.
        </xs:documentation>
    </xs:annotation>
</xs:enumeration>

<xs:enumeration value="GEO_D">
    <xs:annotation>
        <xs:documentation>
            3D Geographic (geocentric) coordinates where the magnitude
            is expressed as an elevation above sea-level.
        </xs:documentation>
        <xs:documentation>
            If applicable, queries against this system should assume a
            default magnitude value or range in the absence of an
            applicable (non-standard) query constraint.  Service
            providers are encouraged to document the such assumptions
            in the resource or capability description.
        </xs:documentation>
        <xs:documentation>
            Semi-major axis and inverse flattening of the
            reference spheroid may need to be assumed; a
            default as defined by the IAU 1976 is recommended.
        </xs:documentation>
    </xs:annotation>
</xs:enumeration>

<xs:enumeration value="MAG">
    <xs:annotation>
        <xs:documentation>
            Geomagnetic coordinates.
        </xs:documentation>
        <xs:documentation>
            See Franz and Harper 2002, Planetary and Space
            Science, vol 50, p. 217.
        </xs:documentation>
        <xs:documentation>
            If applicable, queries against this system should assume a
            default magnitude value or range in the absence of an
            applicable (non-standard) query constraint.  Service
            providers are encouraged to document the such assumptions
            in the resource or capability description.
        </xs:documentation>
    </xs:annotation>
</xs:enumeration>

<xs:enumeration value="GSE">
    <xs:annotation>
        <xs:documentation>
            Geocentric Solar Ecliptic coordinates
        </xs:documentation>
        <xs:documentation>
            See Franz and Harper 2002, Planetary and Space
            Science, vol 50, p. 217.
        </xs:documentation>
    </xs:annotation>
</xs:enumeration>

<xs:enumeration value="GSM">
    <xs:annotation>
        <xs:documentation>
            Geocentric Solar Magnetic coordinates
        </xs:documentation>
```

```xml
            <xs:documentation>
              See Franz and Harper 2002, Planetary and Space
              Science, vol 50, p. 217.
            </xs:documentation>
          </xs:annotation>
        </xs:enumeration>

        <xs:enumeration value="HGC">
          <xs:annotation>
            <xs:documentation>
              Heliographic coordinates (Carrington)
            </xs:documentation>
            <xs:documentation>
              See Thompson 2006, "Coordinate Systems for Solar
              Image Data", A&A., Section 2.2
            </xs:documentation>
          </xs:annotation>
        </xs:enumeration>

        <xs:enumeration value="HGS">
          <xs:annotation>
            <xs:documentation>
              Heliographic coordinates (Stonyhurst)
            </xs:documentation>
            <xs:documentation>
              See Thompson 2006, "Coordinate Systems for Solar
              Image Data", A&A., Section 2.2
            </xs:documentation>
          </xs:annotation>
        </xs:enumeration>

        <xs:enumeration value="HEEQ">
          <xs:annotation>
            <xs:documentation>
              Heliographic Earth Equatorial coordinates
            </xs:documentation>
            <xs:documentation>
              See Franz and Harper 2002, Planetary and Space
              Science, vol 50, p. 217, and
              Thompson 2006, "Coordinate Systems for Solar
              Image Data", A&A., Section 2.1
            </xs:documentation>
          </xs:annotation>
        </xs:enumeration>

        <xs:enumeration value="HRTN">
          <xs:annotation>
            <xs:documentation>
              Heliographic Radial-Tangential-Normal coordinates
            </xs:documentation>
            <xs:documentation>
              See Franz and Harper 2002, Planetary and Space
              Science, vol 50, p. 217.
            </xs:documentation>
          </xs:annotation>
        </xs:enumeration>

        <xs:enumeration value="HPC">
          <xs:annotation>
            <xs:documentation>
              Helioprojective Cartesian coordinates
            </xs:documentation>
            <xs:documentation>
              See Thompson 2006, "Coordinate Systems for Solar
              Image Data", A&A., Section 2.1
            </xs:documentation>
          </xs:annotation>
        </xs:enumeration>

        <xs:enumeration value="HPR">
          <xs:annotation>
            <xs:documentation>
              Helioprojective Polar coordinates
```

```xml
                    </xs:documentation>
                    <xs:documentation>
                        See Thompson 2006, "Coordinate Systems for Solar
                        Image Data", A&A., Section 2.1
                    </xs:documentation>
                </xs:annotation>
            </xs:enumeration>

            <xs:enumeration value="HCC">
                <xs:annotation>
                    <xs:documentation>
                        Heliocentric Cartesian coordinates
                    </xs:documentation>
                    <xs:documentation>
                        See Thompson 2006, "Coordinate Systems for Solar
                        Image Data", A&A., Section 2.1
                    </xs:documentation>
                </xs:annotation>
            </xs:enumeration>

            <xs:enumeration value="HGI">
                <xs:annotation>
                    <xs:documentation>
                        Heliographic Inertial coordinates
                    </xs:documentation>
                    <xs:documentation>
                        See Franz and Harper 2002, Planetary and Space
                        Science, vol 50, p. 217.
                    </xs:documentation>
                </xs:annotation>
            </xs:enumeration>

            <xs:enumeration value="MERCURY_C">
                <xs:annotation>
                    <xs:documentation>
                        Planteocentric coordinates on Mercury
                    </xs:documentation>
                </xs:annotation>
            </xs:enumeration>

            <xs:enumeration value="VENUS_C">
                <xs:annotation>
                    <xs:documentation>
                        Planteocentric coordinates on Venus
                    </xs:documentation>
                </xs:annotation>
            </xs:enumeration>

            <xs:enumeration value="LUNA_C">
                <xs:annotation>
                    <xs:documentation>
                        Selenocentric coordinates (for the Moon)
                    </xs:documentation>
                </xs:annotation>
            </xs:enumeration>

            <xs:enumeration value="MARS_C">
                <xs:annotation>
                    <xs:documentation>
                        Planteocentric coordinates on Mars
                    </xs:documentation>
                </xs:annotation>
            </xs:enumeration>

            <xs:enumeration value="JUPITER_C_III">
                <xs:annotation>
                    <xs:documentation>
                        Planteocentric coordinates on Jupiter, system III
                    </xs:documentation>
                </xs:annotation>
            </xs:enumeration>

            <xs:enumeration value="SATURN_C_III">
```

```xml
                <xs:annotation>
                    <xs:documentation>
                        Planteocentric coordinates on Saturn, system III
                    </xs:documentation>
                </xs:annotation>
            </xs:enumeration>

            <xs:enumeration value="URANUS_C_III">
                <xs:annotation>
                    <xs:documentation>
                        Planteocentric coordinates on Uranus, system III
                    </xs:documentation>
                </xs:annotation>
            </xs:enumeration>

            <xs:enumeration value="NEPTUNE_C_III">
                <xs:annotation>
                    <xs:documentation>
                        Planteocentric coordinates on Neptune, system III
                    </xs:documentation>
                </xs:annotation>
            </xs:enumeration>

            <xs:enumeration value="PLUTO_C">
                <xs:annotation>
                    <xs:documentation>
                        Planteocentric coordinates on Mercury
                    </xs:documentation>
                </xs:annotation>
            </xs:enumeration>

            <xs:enumeration value="MERCURY_G">
                <xs:annotation>
                    <xs:documentation>
                        Planteographic coordinates on Mercury
                    </xs:documentation>
                </xs:annotation>
            </xs:enumeration>

            <xs:enumeration value="VENUS_G">
                <xs:annotation>
                    <xs:documentation>
                        Planteographic coordinates on Venus
                    </xs:documentation>
                </xs:annotation>
            </xs:enumeration>

            <xs:enumeration value="LUNA_G">
                <xs:annotation>
                    <xs:documentation>
                        Selenographic coordinates (for the Moon)
                    </xs:documentation>
                </xs:annotation>
            </xs:enumeration>

            <xs:enumeration value="MARS_G">
                <xs:annotation>
                    <xs:documentation>
                        Planteographic coordinates on Mars
                    </xs:documentation>
                </xs:annotation>
            </xs:enumeration>

            <xs:enumeration value="JUPITER_G_III">
                <xs:annotation>
                    <xs:documentation>
                        Planteographic coordinates on Jupiter, system III
                    </xs:documentation>
                </xs:annotation>
            </xs:enumeration>

            <xs:enumeration value="SATURN_G_III">
                <xs:annotation>
```

```xml
                    <xs:documentation>
                        Planteographic coordinates on Saturn, system III
                    </xs:documentation>
                </xs:annotation>
            </xs:enumeration>

            <xs:enumeration value="URANUS_G_III">
                <xs:annotation>
                    <xs:documentation>
                        Planteographic coordinates on Uranus, system III
                    </xs:documentation>
                </xs:annotation>
            </xs:enumeration>

            <xs:enumeration value="NEPTUNE_G_III">
                <xs:annotation>
                    <xs:documentation>
                        Planteographic coordinates on Neptune, system III
                    </xs:documentation>
                </xs:annotation>
            </xs:enumeration>

            <xs:enumeration value="PLUTO_G">
                <xs:annotation>
                    <xs:documentation>
                        Planteographic coordinates on Mercury
                    </xs:documentation>
                </xs:annotation>
            </xs:enumeration>

            <xs:enumeration value="UNKNOWN">
                <xs:annotation>
                    <xs:documentation>
                        a frame that is either unknown or non-standard
                    </xs:documentation>
                    <xs:documentation>
                        Any available descriptive details should be
                        given (or referenced) in the capability's description.
                    </xs:documentation>
                </xs:annotation>
            </xs:enumeration>

        </xs:restriction>
    </xs:simpleType>

    <xs:complexType name="Query">
        <xs:annotation>
            <xs:documentation>
                A query to be sent to the service
            </xs:documentation>
        </xs:annotation>

        <xs:sequence>
            <xs:element name="pos" type="ssap:PosParam" minOccurs="0">
                <xs:annotation>
                    <xs:documentation>
                        the center position the search cone given in
                        decimal degrees.
                    </xs:documentation>
                </xs:annotation>
            </xs:element>

            <xs:element name="size" type="xs:double" minOccurs="0">
                <xs:annotation>
                    <xs:documentation>
                        the size of the search radius.
                    </xs:documentation>
                </xs:annotation>
            </xs:element>

            <xs:element name="queryDataCmd" type="xs:string" minOccurs="0">
                <xs:annotation>
                    <xs:documentation>
```

```
                    Fully specified test query formatted as an URL
                    argument list in the syntax specified by the SSA standard.
                    The list must exclude the REQUEST argument which is
                    assumed to be set to "queryData".
                </xs:documentation>
                <xs:documentation>
                    This value must be in the form of name=value
                    pairs delimited with apersands (&).  A query
                    may then be formed by appending to the base URL the
                    request argument, "REQUEST=queryData&", followed
                    by the contents of this element.
                </xs:documentation>
            </xs:annotation>
        </xs:element>
    </xs:sequence>
</xs:complexType>

<xs:complexType name="PosParam">
    <xs:annotation>
        <xs:documentation>
            a position in the sky to search.
        </xs:documentation>
    </xs:annotation>

    <xs:sequence>
        <xs:element name="long" type="xs:double">
            <xs:annotation>
                <xs:documentation>
                    The longitude (e.g. Right Ascension) of the center of the
                    search position in decimal degrees.
                </xs:documentation>
            </xs:annotation>
        </xs:element>

        <xs:element name="lat" type="xs:double">
            <xs:annotation>
                <xs:documentation>
                    The latitude (e.g. Declination) of the center of the
                    search position in decimal degrees.
                </xs:documentation>
            </xs:annotation>
        </xs:element>

        <xs:element name="refframe" type="xs:token" minOccurs="0">
            <xs:annotation>
                <xs:documentation>
                    the coordinate system reference frame name indicating
                    the frame to assume for the given position.   If not
                    provided, ICRS is assumed.
                </xs:documentation>
            </xs:annotation>
        </xs:element>
    </xs:sequence>
</xs:complexType>

<!--
  -  a separate type for pre-v1.0 compliant spectral services.
  -->
<xs:complexType name="ProtoSpectralAccess">
    <xs:annotation>
        <xs:documentation>
            The capabilities of an proto-SSA service implementation.
            Clients may assume a particular interface for this type of
            service based on historical convention; however, no
            guarantees are made that the service is compliant with any
            IVOA standard.
        </xs:documentation>
        <xs:documentation>
            This capability is for spectral access services developed
            prior to the completion of the SSA standard and,
            therefore, are not compliant with that standard.
        </xs:documentation>
    </xs:annotation>
```

```xml
<xs:complexContent>
    <xs:extension base="ssap:SSACapRestriction">
        <xs:sequence>

            <xs:element name="dataSource" type="ssap:DataSource"
                        minOccurs="1" maxOccurs="unbounded">
                <xs:annotation>
                    <xs:documentation>
                        The category specifying where the data originally
                        came from.
                    </xs:documentation>
                    <xs:documentation>
                        Allowed values are "survey", "pointed", "custom",
                        "theory", "artificial"
                    </xs:documentation>
                </xs:annotation>
            </xs:element>

            <xs:element name="creationType" type="ssap:CreationType"
                        minOccurs="1" maxOccurs="unbounded">
                <xs:annotation>
                    <xs:documentation>
                        The category that describes the process used to
                        produce the dataset.
                    </xs:documentation>
                    <xs:documentation>
                        Typically this describes only the processing
                        performed by the data service, but it could
                        describe some additional earlier processing as
                        well, e.g., if data is partially precomputed.
                    </xs:documentation>
                    <xs:documentation>
                        Allowed values are "archival", "cutout", "filtered",
                        "mosaic", "projection", "spectralExtraction",
                        "catalogExtraction"
                    </xs:documentation>
                </xs:annotation>
            </xs:element>

            <xs:element name="maxSearchRadius" type="xs:double"
                        minOccurs="0" maxOccurs="1">
                <xs:annotation>
                    <xs:documentation>
                        The largest search radius, in degrees, that will be
                        accepted by the service without returning an error
                        condition.  Not providing this element or
                        specifying a value of 180 indicates that there
                        is no restriction.
                    </xs:documentation>
                    <xs:documentation>
                        Not providing a value is the prefered way to
                        indicate that there is no restriction.
                    </xs:documentation>
                </xs:annotation>
            </xs:element>

            <xs:element name="maxRecords" type="xs:positiveInteger">
                <xs:annotation>
                    <xs:documentation>
                        The hard limit on the largest number of records that
                        the query operation will return in a single response
                    </xs:documentation>
                </xs:annotation>
            </xs:element>

            <xs:element name="defaultMaxRecords" type="xs:positiveInteger">
                <xs:annotation>
                    <xs:documentation>
                        The largest number of records that the service will
                        return when the MAXREC parameter not specified
                        in the query input.
                    </xs:documentation>
```

```
                </xs:annotation>
            </xs:element>

            <xs:element name="maxAperture" type="xs:double"
                        minOccurs="0">
                <xs:annotation>
                    <xs:documentation>
                        The largest aperture diameter that can be supported
                        upon request via the APERTURE input parameter by a
                        service that supports the spectral extraction
                        creation method.  A value of 360 or not providing
                        a value means there is no theoretical limit.
                    </xs:documentation>
                    <xs:documentation>
                        Not providing a value is the preferred way to
                        indicate that there is no limit.
                    </xs:documentation>
                </xs:annotation>
            </xs:element>

            <xs:element name="maxFileSize" type="xs:int"
                        minOccurs="0" maxOccurs="1">
                <xs:annotation>
                    <xs:documentation>
                        The maximum image file size in bytes.
                    </xs:documentation>
                </xs:annotation>
            </xs:element>

            <xs:element name="testQuery" type="ssap:Query"
                        minOccurs="0" maxOccurs="1">
                <xs:annotation>
                    <xs:documentation>
                        a set of query parameters that is expected to
                        produce at least one matched record which can be
                        used to test the service.
                    </xs:documentation>
                </xs:annotation>
            </xs:element>

        </xs:sequence>
      </xs:extension>
    </xs:complexContent>
  </xs:complexType>

</xs:schema>
```

## A.4. The SLAP XML Schema

**The Complete SLA Schema**

```
<?xml version="1.0" encoding="UTF-8"?>
<xs:schema xmlns:xs="http://www.w3.org/2001/XMLSchema"
           xmlns:vr="http://www.ivoa.net/xml/VOResource/v1.0"
           xmlns:slap="http://www.ivoa.net/xml/SLAP/v1.0"
           xmlns:vm="http://www.ivoa.net/xml/VOMetadata/v0.1"
           targetNamespace="http://www.ivoa.net/xml/SLAP/v1.0"
           elementFormDefault="unqualified" attributeFormDefault="unqualified"
           version="1.0">

    <!--
         First version 09/09/09
         Authors: Jesus Salgado/Aurelien Stebe (ESAVO)
      -->
    <xs:annotation>
        <xs:appinfo>
           <vm:schemaName>SLAP</vm:schemaName>
           <vm:schemaPrefix>xs</vm:schemaPrefix>
           <vm:targetPrefix>slap</vm:targetPrefix>
        </xs:appinfo>
        <xs:documentation>
```

```
              XML Schema used to describe the capabilities of a service instance
              conforming to the Simple Line Access Protocol (SLAP).
         </xs:documentation>
      </xs:annotation>

   <xs:import namespace="http://www.ivoa.net/xml/VOResource/v1.0"
              schemaLocation="http://www.ivoa.net/xml/VOResource/v1.0"/>

   <!-- Set the Capability standardID to indicate the SLAP protocol. -->
   <xs:complexType name="SLAPCapRestriction" abstract="true">
      <xs:annotation>
         <xs:documentation>
            An abstract capability that fixes the standardID to the
            IVOA ID for the SLAP standard.
         </xs:documentation>
         <xs:documentation>
            See vr:Capability for documentation on inherited children.
         </xs:documentation>
      </xs:annotation>

      <xs:complexContent>
         <xs:restriction base="vr:Capability">
            <xs:sequence>
               <xs:element name="validationLevel" type="vr:Validation"
                           minOccurs="0" maxOccurs="unbounded"/>
               <xs:element name="description" type="xs:token"
                           minOccurs="0"/>
               <xs:element name="interface" type="vr:Interface"
                           minOccurs="0" maxOccurs="unbounded"/>
            </xs:sequence>
            <xs:attribute name="standardID" type="vr:IdentifierURI"
                          use="required" fixed="ivo://ivoa.net/std/SLAP"/>
         </xs:restriction>
      </xs:complexContent>
   </xs:complexType>

   <!-- SLAP Capabilities -->
   <xs:complexType name="SimpleLineAccess">
      <xs:annotation>
         <xs:documentation>
            The capabilities of an SLAP service implementation.
         </xs:documentation>
      </xs:annotation>

      <xs:complexContent>
         <xs:extension base="slap:SLAPCapRestriction">
            <xs:sequence>

               <xs:element name="complianceLevel" type="slap:ComplianceLevel">
                  <xs:annotation>
                     <xs:documentation>
                        The category indicating the level to which this
                        service instance complies with the SLAP standard.
                     </xs:documentation>
                     <xs:documentation>
                        Allowed values are "minimal" and "full".
                        See definitions of allowed values for details.
                     </xs:documentation>
                  </xs:annotation>
               </xs:element>

               <xs:element name="dataSource" type="slap:DataSource">
                  <xs:annotation>
                     <xs:documentation>
                        The category specifying where the data accessed by
                        the service originally came from.
                     </xs:documentation>
                     <xs:documentation>
                        Allowed values are "observational/astrophysical",
                        "observational/laboratory", "theoretical"
                     </xs:documentation>
                  </xs:annotation>
               </xs:element>
```

```xml
                <xs:element name="maxRecords" type="xs:positiveInteger"
                        minOccurs="0" maxOccurs="1">
                    <xs:annotation>
                        <xs:documentation>
                            The hard limit on the largest number of records that
                            the query operation will return in a single response.
                            Not providing this value means that there is no
                            effective limit.
                        </xs:documentation>
                        <xs:documentation>
                            This does not refer to the total number of spectra in
                            the archive but rather maximum number of records the
                            service is capable of returning.  A limit that is
                            greater than the number of spectra available in the
                            archive is equivalent to their being no effective
                            limit.  (See RM, Hanisch 2007.)
                        </xs:documentation>
                    </xs:annotation>
                </xs:element>

                <xs:element name="testQuery" type="slap:Query"
                        minOccurs="0" maxOccurs="1">
                    <xs:annotation>
                        <xs:documentation>
                            A set of queryData parameters that is expected to
                            produce at least one matched record which can be
                            used to test the service.
                        </xs:documentation>
                        <xs:documentation>
                            The value should include all parameters required
                            for the test query but should exclude the baseURL
                            and the REQUEST parameter.
                        </xs:documentation>
                    </xs:annotation>
                </xs:element>

            </xs:sequence>
        </xs:extension>
    </xs:complexContent>
</xs:complexType>

<xs:simpleType name="ComplianceLevel">
    <xs:annotation>
        <xs:documentation>
            The allowed values for indicating the level at which a service
            instance complies with the SLAP standard.
        </xs:documentation>
    </xs:annotation>

    <xs:restriction base="xs:token">
        <xs:enumeration value="minimal">
            <xs:annotation>
                <xs:documentation>
                    The service supports all of the capabilities and features
                    of the SLAP protocol identified as "must" in the
                    specification.
                </xs:documentation>
                <xs:documentation>
                    In brief, this includes:
                        * implementing the GET interface,
                        * WAVELENGTH, REQUEST for input query,
                        * ssldm:Line.wavelength.value and ssldm:Line.title for output
                        fields
                        * supports the "FORMAT=METADATA" metadata query.
                </xs:documentation>
            </xs:annotation>
        </xs:enumeration>

        <xs:enumeration value="full">
            <xs:annotation>
                <xs:documentation>
                    The service supports, at a minimum, all of the capabilities
```

```
                            and features of the SLAP protocol identified as "must" or
                            "should" in the specification.
                        </xs:documentation>
                    </xs:annotation>
                </xs:enumeration>
            </xs:restriction>
        </xs:simpleType>

        <xs:simpleType name="DataSource">
            <xs:annotation>
                <xs:documentation>
                    The defined categories that specify where the line data
                    originally came from, i.e., the type of data collections to
                    which the service provides access.
                </xs:documentation>
            </xs:annotation>

            <xs:restriction base="xs:token">
                <xs:enumeration value="observational/astrophysical">
                    <xs:annotation>
                        <xs:documentation>
                            Lines observed and identified in real spectra of
                            astrophysical observations by different
                            instrument/projects
                        </xs:documentation>
                    </xs:annotation>
                </xs:enumeration>

                <xs:enumeration value="observational/laboratory">
                    <xs:annotation>
                        <xs:documentation>
                            Lines observed and identified in real spectra of
                            laboratory measurements
                        </xs:documentation>
                    </xs:annotation>
                </xs:enumeration>

                <xs:enumeration value="theoretical">
                    <xs:annotation>
                        <xs:documentation>
                            Servers containing theoretical spectral lines
                        </xs:documentation>
                    </xs:annotation>
                </xs:enumeration>

            </xs:restriction>
        </xs:simpleType>

        <xs:complexType name="Query">
            <xs:annotation>
                <xs:documentation>
                    A query to be sent to the service, e.g., a test query.
                </xs:documentation>
            </xs:annotation>

            <xs:sequence>
                <xs:element name="wavelength" type="slap:WavelengthRange" minOccurs="0">
                    <xs:annotation>
                        <xs:documentation>
                            Spectral range in meters to be used to constrain the query
                            of spectral lines.
                        </xs:documentation>
                    </xs:annotation>
                </xs:element>

                <xs:element name="queryDataCmd" type="xs:string" minOccurs="0">
                    <xs:annotation>
                        <xs:documentation>
                            Fully specified queryData test query formatted as an URL
                            argument list in the syntax specified by the SLAP standard.
                            The list must exclude the REQUEST argument which is
                            assumed to be set to "queryData".  VERSION may be
                            included if the test query applies to a specific version
```

```
                         of the service protocol.
                    </xs:documentation>
                    <xs:documentation>
                         If queryDataCmd is used to form a query, the default
                         value of WAVELENGTH specified above is not
                         used; if the test query requires WAVELENGTH it
                         should be included directly in queryDataCmd.
                    </xs:documentation>
                    <xs:documentation>
                         This value must be a string in the form of name=value
                         pairs delimited with ampersands (&).  A query may
                         then be formed by appending to the baseURL the request
                         argument, "REQUEST=queryData&", followed by the
                         contents of this element.
                    </xs:documentation>
                </xs:annotation>
            </xs:element>
        </xs:sequence>
    </xs:complexType>

    <xs:complexType name="WavelengthRange">
        <xs:annotation>
            <xs:documentation>
                Spectral range in meters to be used to constrain the query
                of spectral lines
            </xs:documentation>
        </xs:annotation>

        <xs:sequence>
            <xs:element name="minWavelength" type="xs:double" minOccurs="0">
                <xs:annotation>
                    <xs:documentation>
                         Minimum wavelength in meters to be used to constrain the query
                         of spectral lines
                    </xs:documentation>
                </xs:annotation>
            </xs:element>

            <xs:element name="maxWavelength" type="xs:double" minOccurs="0">
                <xs:annotation>
                    <xs:documentation>
                         Maximum wavelength in meters to be used to constrain the query
                         of spectral lines
                    </xs:documentation>
                </xs:annotation>
            </xs:element>
        </xs:sequence>
    </xs:complexType>

</xs:schema>
```

# Appendix B: Supporting Multiple Versions of DAL Protocols

*Note: this section is non-normative.*

It is possible for a VOResource-encoded resource description to indicate support for multiple versions of standard service. This is described generally in Section 2.2.2 ("The Service Data Model") of the VOResource specification [VOR]. In that section, the specification says that a `<capability>` element can contain multiple `<interface>` elements, each describing a different version. In this appendix, we illustrate how this can be applied to the DAL services covered by this (SimpleDALRegExt) specification.

We start by noting that the `standardID` values for each of the DAL protocols described in this document refer to the standards generally, without reference to the particular version. For example, the IVOA identifier for the Simple Cone Search protocol is `ivo://ivoa.net/std/ConeSearch`. Thus a `<capability>` element can logically describe support for any version or multiple versions of the standard DAL protocol as long as the extension schema for that protocol is same for all of the versions.

Here is an example a service that supports two versions of the SSA protocol:

**Example**

Supporting multiple versions of the SSA standard. Below shows just the `<capability>` element of a description of an SSA service resource.

```
<capability xsi:type="ssap:SimpleSpectralAccess" standardID="ivo://ivoa.net/std/SSA">

    <description>
      This service supports the SSA protocol standard.  Currently, there is
      support for both 1.04 and 1.1 versions of the spec.  A browser-based
      front-end is also available.
    </description>

    <!--
      - this gives the base URL for the standard GET interface, version 1.1
      -->
    <interface xsi:type="vs:ParamHTTP" role="std" version="1.1">
       <accessURL use="base">
         http://adil.ncsa.uiuc.edu/cgi-bin/vossa11
       </accessURL>
    </interface>

    <!--
      - this gives the base URL for the standard GET interface, version 1.04
      -->
    <interface xsi:type="vs:ParamHTTP" role="std" version="1.04">
       <accessURL use="base">
         http://adil.ncsa.uiuc.edu/cgi-bin/vossa
       </accessURL>
    </interface>

    <!--
      - Here's an interactive version.  Since it's just a front-end to
      - the GET service, it shares the same capability metadata.
      - This element does not have a version attribute nor is role="std"
      - set, since it is not part of the standard.
      -->
    <interface xsi:type="vr:WebBrowser">
       <accessURL> http://adil.ncsa.uiuc.edu/ws/vossa </accessURL>
    </interface>

    <!--
      - This starts the SSA-specific metadata
      -->
    <complianceLevel>full</complianceLevel>
    <dataSource>pointed</dataSource>
    <creationType>cutout</creationType>
    <maxSearchRadius>10</maxSearchRadius>
    <maxRecords>10000</maxRecords>
    <defaultMaxRecords>500</defaultMaxRecords>
    <maxAperture>3600</maxAperture>

</capability>
```

In the above example, two different SSA service endpoints are supported, each compliant with a different version of the SSA standard. The `version` attribute on the `<interface>` elements indicate which version the endpoint supports. Note that if a `version` attribute is not provided, it defaults to "1.0". The `role="std"` distinguishes standard-complying interfaces from custom ones; in the example above, an endpoint to a browser-based front-end is provided as well.

A consumer that wishes to engage a standard DAL service, should extract interface descriptions where `role="std"` is set. From of those, the consumer should inspect the `version` attribute to select a version that it supports.

Often, the difference between versions is of little consequence to a consumer, so having to choose between several standardized versions is an unneeded complication. Thus, it is recommended that publishers list the most preferred version for users (typically the latest) first among the list of interfaces; consequently, consumers that do not care to attempt to interpret the version attribute should choose the first in the list. If the service query fails (because the consumer's assumptions about the service protocol are not applicable to that version of the service), the consumer may try the next one in the list.

# Appendix C: Change History

**Changes since PR-v1.0 20130911:**

- none other than date and status.

**Changes from PR-v1.0 20121116**

- for SSA's `creationType`, changed `specialExtraction` to `spectralExtraction`.
- corrected Creation Type reference to section in SSA doc.
- made `long` and `lat` elements in `ssap:PosParam` required.
- incremented SSA schema version to 1.1 in namespace.
- refresh App. A from official schemas
- fixed typos ("lRCS" and value type for `maxFileSize)`
- noted that the `<long>` and `<lat>` values within the `sia:SkySize` type are given in degrees.
- Fixed documentation of SIA's `sia:Query` type in the schema.

**Changes from PR-v1.0 20120517**

- The namespace URIs given in Sections 3.1.1, 3.2.1, 3.3.1, and 3.4.1 were updated to match that specified in the XSDs (i.e. to include a "v" preceding the version field).
- Several capability metadata with types `xs:int` and `xs:float` were changed to `xs:positiveInterger` `xs:double` to allow for larger/more precise numbers.
- Capability metadata that indicated maximum allowed values (e.g. `<maxRecords>`, `<maxImageSize>`, etc.) were made optional to avoid large, meaningless numbers from being provided. Now not specifying a value is the preferred way to indicate that no upper limit applies.
- Semantic definition of `<sia:maxImageExtent>` clarified to differentiate it from `<sia:maxQueryRegionSize>`
- The type for `<sia:maxImageSize>` was changed to `xs:positiveInteger`, a single number that represents the length of a side in pixels. The `sia:ImageSize` type (no longer needed) was dropped.
- The version field in the SIA namespace was incremented to 1.1 due to the non-backward-compatible change to `<sia:maxImageSize>`
- various typos and grammatical errors corrected.

**Changes from WD-v1.0 20110921:**

- Now recommend `ssap` as prefix; changed all occurances of `ssa` in text and schema.
- added `<supportedFrame>` to `ssap:SimpleSpectralAccess`
- removed import of VODataService schema from SIA, SSA, and Conesearch schemas.
- change base type of controlled vocab types from `xs:string` to `xs:token` for consistancy with VOResource.

# References


**[RFC 2119]**
Bradner, S. 1997. *Key words for use in RFCs to Indicate Requirement Levels*, IETF RFC 2119, `http://www.ietf.org/rfc/rfc2119.txt`

**[schema]**
Fallside, David C., Walmsley, Priscilla (editors) 2004, *XML Schema Part 0: Primer Second Edition*, W3C Recommendation 28 October 2004, `http://www.w3.org/TR/xmlschema-0/`

**[Arch]**
Arviset, Christophe and the IVOA Tehcnical Coordination Group 2010, *The IVOA in 2010: Technical Assessment and Roadmap*, v1.0, IVOA Note, *in preparation*.

**[SCS]**
Williams, R., Hanisch, R., Szalay, A., and Plante, R. 2008, *Simple Cone Search Version 1.03*, IVOA Recommendation, 22 February 2008, `http://www.ivoa.net/Documents/REC/DAL/ConeSearch-20080222.html`.



**[SIA]**

Tody, D. and Plante, R. 2004, *Simple Image Access Specification Version 1.0*, IVOA Working Draft, `http://www.ivoa.net/Documents/WD/SIA/sia-20040524.html`.

**[SSA]**

Tody, D., Dolensky, M., McDowell, J., Bonnarel, F., Budavari, T., Busko, I., Micol, A., Osuna, P, Salgado, J., Skoda, P., Thompson, R., and Valdes, F. 2008, *Simple Spectral Access Protocol Version 1.04*, IVOA Recommendation, 01 Febrary 2008, `http://www.ivoa.net/Documents/cover/SSA-20080201.html`.

**[SLA]**

Salgado, J., Osuna, P., Guainazzi, M., Barbarisi, I., Dubernet, M., and Tody. D. 2010, *Simple Line Access Protocol Version 1.0*, IVOA Recommendation, 9 December 2010, `http://www.ivoa.net/Documents/SLAP/20101209/`.

**[RI]**

Benson, Kevin, Plante, Ray, Auden, Elizabeth, Graham, Matthew, Greene, Gretchen, Hill, Martin, Linde, Tony, Morris, Dave, O'Mullane, Wil, Rixon, Guy, Andrews, Kona 2008, *IVOA Registry Interfaces*, v1.02, IVOA Recommendation `http://www.ivoa.net/Documents/latest/ResourceInterface.html`

**[RM]**

Hanisch, Robert (ed.) 2004, *Resource Metadata for the Virtual Observatory, Version 1.12*, IVOA Recommendation, `http://www.ivoa.net/Documents/REC/ResMetadata/RM-20040426.htm`

**[VOR]**

Plante, R., Benson, K., Graham, M., Greene, G., Harrison, P., Lemson, G., Linde, T., Rixon, G., Stébé, A. 2008, *VOResource: an XML Encoding Schema for Resource Metadata*, v1.03, IVOA Recommendation, `http://www.ivoa.net/Documents/REC/ReR/VOResource-20080222.html`

**[VDS]**

Plante, R., Stébé, A. Benson, K., Dowler, P. Graham, M., Greene, G., Harrison, P., Lemson, G., Linde, T., Rixon, G., Stébé, A. 2009, *VODataService: a VOResource Schema Extension for Describing Collections and Services*, v1.1, IVOA Recommendation, `http://www.ivoa.net/Documents/VODataService/20101202/`

**[STC]**

Rots, Arnold 2007, *Space-Time Coordinate Metadata for the Virtual Observatory*, v1.33, IVOA Recommendation, `href="http://www.ivoa.net/Documents/REC/STC/STC-20071030.html">`